\documentclass[a4paper]{article}
\usepackage{amssymb,latexsym,bm}
\usepackage{graphicx,color}
\usepackage[numbers,sort&compress]{natbib}
\bibpunct[, ]{[}{]}{,}{n}{,}{,}
\makeatletter
\def\NAT@def@citea{\def\@citea{\NAT@separator}}
\makeatother

\def\bra{\langle}
\def\ket{\rangle}
\def\p{\partial}

\def\beq{\begin{equation}}
\def\eeq{\end{equation}}
\def\la{\label}

\def\r#1{(\ref{#1})}
\newcommand{\mylab}[3]{\raisebox{#2}[0mm][0mm]{%
\makebox[0mm][l]{\hspace*{#1}{#3}}}}%

\def\uvec{\mbox{\boldmath $u$}}
\def\xvec{\mbox{\boldmath $x$}}

\def\bvec{\mbox{\boldmath $b$}}
\def\valpha{{\bm \alpha}}
\def\vdelta{{\bm \delta}}
\def\matI{\mbox{\rm\bf I}}
\def\matA{\mbox{\rm\bf A}}
\def\sref{{\mbox{\small\it ref}}}
\def\spacce#1{\hskip #1pt}
\def\drawline#1#2{\raise 2.5pt\vbox{\hrule width #1pt height #2pt}}
\def\solid{\drawline{24}{.5}\nobreak}
\def\shortsolid{\drawline{8}{.5}\nobreak}

\def\bdashli{\hbox{\drawline{5}{.5}\spacce{2}}}
\def\bdash{\hbox{\drawline{5.8}{.5}\spacce{2}}}

\def\dashed{\bdash\bdash\bdash\nobreak}
\def\dashedshort{\bdashli\bdashli\nobreak}

\def\chndot{\hbox%
{\drawline{4.6}{.5}\spacce{2}\drawline{1}{.5}\spacce{2}\drawline{4.6}{.5}\spacce{2}\drawline{1}{.5}\spacce{2}\drawline{4.6}{.5}}\nobreak }
\def\circle{$\circ$\nobreak }

\def\trian{\raise 1.25pt\hbox{$\scriptstyle\triangle$}\nobreak}

\def\dtrian{\raise 1.25pt\hbox%
{$\scriptscriptstyle\bigtriangledown$}\nobreak}
\def\rtrian{\raise 1.25pt\hbox{$\triangleright$}\nobreak}
\def\ltrian{\raise 1.25pt\hbox{$\triangleleft$}\nobreak}

\def\squar{\raise 1.25pt\hbox{$\scriptstyle\Box$}\nobreak}

\def\diamon{\raise 1.25pt\hbox{$\scriptstyle\diamond$}\nobreak}

\newcommand{\soliddtrian}{$\blacktriangledown$\nobreak}

\def\linedtri1{\hbox{\bdash\hspace{-1.6mm}$\bigtriangleup$\hspace{-0.8mm}\bdash}\nobreak}
\def\soliddtrian1{$\blacktriangledown$\nobreak}
\def\solidrtrian2{$\blacktriangleright$\nobreak}
\def\solidltrian3{$\blacktriangleleft$\nobreak}
\begin{document}

\title{Monte-Carlo Science}

\author{%
Javier Jim\'enez\thanks{CONTACT Javier Jim\'enez. Email: javier.jimenezs@upm.es}\\
School of Aeronautics, Universidad Polit\'ecnica Madrid, 28040 Madrid, Spain
}

\maketitle
\begin{abstract}
This paper explores how far the scientific discovery process can be automated. Using the
identification of causally significant flow structures in two-dimensional turbulence as an
example, it probes how far the usual procedure of planning experiments to test hypotheses
can be substituted by `blind' randomised experiments, and notes that the increased
efficiency of computers is beginning to make such a `Monte-Carlo' approach practical in
fluid mechanics. The process of data generation, classification and model creation is
described in some detail, stressing the importance of validation and verification. Although
the purpose of the paper is to explore the procedure, rather than to model two-dimensional
turbulence, it is encouraging that the Monte Carlo process naturally leads to the
consideration of vortex dipoles as building blocks of the flow, on a par with the more
conventional individual vortex cores. Although not completely novel, this `spontaneous'
discovery supports the claim that an important advantage of randomised experiments is to
bypass researcher prejudice and alleviate paradigm lock. It is finally noted that the method
can be extended to three-dimensional flows in practical times.
\end{abstract}


\section{Introduction}

The business of science is to search for laws that can be used to make predictions. As such,
the classical point of view is that a prerequisite for science is the determination of
causal relations ({\em Does} A cause B?), preferably complemented by mechanisms ({\em How} A
causes B?). Both are important. Causality without mechanisms is unlikely to lead to
quantitative predictions, and mechanisms without causality have a high probability of being
wrong. Both have to be testable, preferably using different data from their training sets.
For example, the ancient Greeks invented causes and mechanisms involving gods and chariots
for the alternation of night and day. They fitted the known facts well, but we know today
that they are hard to generalise.

There are two meanings of the word `science'. The first one is a systematic method for
discovering how things work \cite{poincare08}, and the second is the resulting body of
knowledge about how they work \cite{kuhn70}. Depending of who you ask, one aspect is
considered more important than the other but, without taking sides in this dichotomy, this
paper deals mostly with methodological aspects. Its contribution to the body of knowledge
of fluid mechanics is not intended to be major.

In particular, we explore how to determine causality in physical systems, and 
whether recent developments provide new ways of doing so. To fix ideas, we illustrate
our argument with the study of a particularly flow -- two-dimensional decaying homogeneous
turbulence -- about which the general feeling is that most things are understood. As such, we
will be able to test our conclusions against accepted wisdom, although this may lead some readers to
think that the discussion is uninteresting, because nothing new is likely to be discovered.
We will see that this is not quite true. Some new aspects can still be unearthed, or at
least rediscovered, and the method for doing so is interesting because it would
not have been practical a decade ago.

Similarly, other readers could be excused for wondering why a paper dealing with apparently
philosophical questions about causality, and centred on an unfashionable flow, should be
included in a special issue on the use of artificial intelligence (AI) in fluid mechanics.
In a sense, they would be right. Artificial intelligence, as the term is mostly used at the
moment, is a collection of techniques for the exploitation of large quantities of data. In
terms of the scientific method, this corresponds to the generation of empirical knowledge,
which is only one of several steps in scientific exploration. What interests us here is the
full process of data exploitation, from their generation to their final incorporation into
hypotheses. It will become clear that AI is only occasionally useful in the process, mainly
as a tool, but that computers are indispensable at most stages of it.

There are several distinctions to be made before we begin our argument. The first one has to
do with how AI is supposed to work \cite{nilsson1998}, which is traditionally divided into
{\em symbolic} \cite{New:Sim:76} and {\em sub-symbolic} \cite{Brooks:90}. Symbolic AI is the
classical kind, which manipulates symbols representing real-world variables according to
rules set by the programmer, typically embodied into an `expert system'. For example,
assuming some personal definition of vortices, AI can be used to encapsulate this knowledge
into rules to identify and isolate them. On the contrary, sub-symbolic AI is not interested
in rules, but in algorithms to do things. For example, given enough snapshots of a flow in
which vortices have been identified (maybe by a pre-existing expert system) we can train a
neural network to distinguish them from vorticity sheets. The result of sub-symbolic AI is
not the rule, but the algorithm, and it does not imply that a rule exists. After training
our system, we may not know (or care) what a vortex is, or what distinguishes it from a
sheet, but we may have a faster way of distinguishing one from the other than what would
have been possible using only pre-ordained physics-based rules. Copernicus and the Greeks
had symbolic representations of the day-night cycle, although with very different ideas of
the rules involved. Most other living beings, which can distinguish night from day and
usually predict quite accurately dawn and dusk, are (probably) sub-symbolic. The classical
scientific method, including causality, is firmly on the symbolic side of the divide: we are
not only interested in the result, but also in the rule. But there is a small but growing
body of scientists and engineers, who feel that data and a properly trained
algorithm are all the information required about Nature, and that no further rule is
necessary, as for example discussed in \cite{Cove:16,Succi:18}.

Engineering and medicine have always been at least partly empirical and data-driven, but
biochemistry was probably the first subject to reach the sub-symbolic level in modern times,
after fast methods of DNA synthesis became available in the 1980s. Much of the modern
discussion on raw empiricism relates to it \cite{Voit:19}, and so did the first
article claiming to describe a `robot scientist' \cite{Adam:09}. The availability of enough
data to even consider sub-symbolic fluid mechanics was only recently made possible by the
numerical simulations of the 1990's, which gave us for the first time the feeling that `we
knew everything', and that any question that could be posed to a computer would eventually
be answered \cite{bren:eld:fre:19,jimploff20}. Of course, even without practical
considerations of cost, this did not mean that all questions were answered, because they
first had to be put to the computer by a researcher.

In this paper we discuss to what extent this last roadblock can be removed. The new enabling
technology is the increased speed and memory capacity of computers, which can do in minutes
what used to take days a decade ago. We will argue that this enables a new way of asking
questions, not based on plausible hypothesis, but randomly, in the hope that some of them
might turn out to be interesting. This `Monte Carlo' procedure does not avoid the necessity
of answering the questions, which can presumably also be done by the computer, nor of
evaluating how `interesting' are the answers, which most probably computers cannot yet do.
It neither points towards a future of `human-less' research, but to a new level of
partnership between humans and computers. Similar steps have been taken before: we no longer
dig canals or throw spears by hand, except as a sport, nor do we do arithmetic with pencil
and paper, or integrate differential equations using special functions. Most of these
human-machine symbioses are usually considered beneficial, although many created their own
disruptions when they were introduced. There is no reason to believe that this time will be
different, but it is fair to question which the new advantages and difficulties will be.
 
Relinquishing control of the questions to be asked is not an altogether new experience
to anybody who has trained graduate students, mentored postdocs, or managed an
industrial or academic research group. Any such person knows the feeling
that, at some point, the research is no longer yours, although most of us console ourselves
by arguing that we are training our peers and that, in the end, the overall direction is
set be us. Monte Carlo science lacks both of these (probably spurious) consolations. This
may be its main advantage. It is inevitable that our students or subordinates share some of
our ideas and, most probably, some of our prejudices. It is often argued that
researcher prejudice is the main roadblock to qualitative scientific advances, and that
`paradigm shifts', always hard to come by, are delayed by it \cite{kuhn70}. The main
advantage to be gained from a Monte-Carlo questioning algorithm is probably its lack of
prejudice. If we can avoid transferring our biases to it, such an algorithm would act as an
efficient, {\em unprejudiced}, although probably not yet very smart, scientific assistant.

\begin{figure}
\centering
\framebox[9cm]{\begin{minipage}{8.8cm}

\begin{enumerate}\renewcommand{\labelenumi}{\rm S\arabic{enumi}:}
\item\la{step:observe} Make observations, ask questions about them, and gather information.

\item\la{step:model} Form hypotheses to describe what has been observed, and make predictions. 

\item\la{step:test} Test the predictions against known or new observations, and accept,
reject, or modify the hypothesis accordingly.
\end{enumerate}
\end{minipage}}
\caption{The scientific method}\la{alg:method}
\end{figure}

There is a second distinction that, although related to the previous one, is independent
from it. We have mentioned the importance of causes, but research is not always geared
towards them. The classical description of the scientific method is summarised in figure
\ref{alg:method}, which emphasised its iterative character. The causal relations mentioned
above are encapsulated in the modelling and testing steps, S2 and S3, and emphasis is often
put on them rather than on the data-gathering step S1. In fact, the research loop is often
started in S2, and only later are hypotheses tested against observations, as in
S2--S1--S3. The implied relation is not always causal. Even in data-driven research, where
observations precede hypotheses, the argument is often that ``if A precedes B'', A is likely
to be the cause of B, although it is generally understood that correlation and causation are
not equivalent. A classical example is the observation of night and day mentioned above. The
correlation between earlier days and later nights is perfect, but it does not imply
causation \cite{Makie:1974}.

On the other hand, the concept of causality is not without problems, and it has been argued
that it is indistinguishable from initial conditions in systems described by differential
equations, such as fluid mechanics \cite{russ:12}. While this is true, and we could frame
our goal as a quest to classify initial conditions in terms of their outcome, the result may
not be very informative. We will restrict ourselves here to a shorter-term definition of the
type of: ``The falling tree causes the crashing noise", even if we know that the fall of the
tree has its own reasons for happening, and that such intermediate causes depend on the time
horizon that we impose on them. Such conclusions require something beyond correlations, and
imply an active intervention of the observer in the generation of data to interrupt the flow
of causality from the notional `original' cause into something more immediate. This is
particularly important in turbulence and other chaotic systems, especially if we want to
retain some control authority in spite of the loss of system memory.

Simplifying a lot, the difference between the two points of view, which mostly affects 
how and when step S1 is undertaken in figure \ref{alg:method}, can be
summarised as that 
%
%
{\em observational} science searches for correlations between observations,
and generalises them into models. It is the science of
prediction, and is sub-symbolic at heart.

On the contrary, {\em interventional} science relies on the results of experiments in which some
condition is changed by the observer, generally in the hope of uncovering `causal' (i.e., if
this, then that), or `counterfactual' relations (if not this, then not that)
\cite{popper63}. It is the science of control, and usually at least aims for 
symbolic intelligence.

%

The present paper is oriented towards control. In the example above, we care about the
falling tree because we might want to prevent it from making
noise. A prediction paper would worry more about establishing correlations between tree
health and noise levels, without being especially interested in causality, and would 
be very different from the present one.
    
The rest of the paper is structured as follows. Section \S\ref{sec:tur2d} describes how the
Monte Carlo ideas in this introduction can be applied to the particular problem of
two-dimensional turbulence, and \S\ref{sec:physics} briefly discusses the physics that can
be learned from them. The paper closes in \S\ref{sec:conc} with a discussion of the
connections that can be established between the scientific method and modern data analysis,
in the light of the experience of the previous two sections. Two appendices collect
implementation details required by those interested in reproducing the experiments in
\S\ref{sec:tur2d}. Appendix \ref{sec:tnorm} describes how the temporal horizon of our causal
perturbations is chosen, and appendix \ref{sec:optcomp} explains how to impose perturbations
on vector properties, such as the velocity.

\section{Causality in two-dimensional turbulence: a case study}\la{sec:tur2d}\la{sec:data}

Two-dimensional turbulence is an old problem in fluid mechanics, often discussed in relation
with geophysical flows \cite{maltrud91}. It also appears naturally in highly stratified
situations, which tend to collapse eddies into almost two-dimensional pancakes
\cite{Vor:Afa:Fil:91}, and has more recently been studied in Bose--Einstein condensates
\cite{NeelyEtal:2010}. It fundamentally differs from three-dimensional turbulence in that
the absence of vortex stretching inhibits the generation of velocity gradients, and
therefore the usual energy cascade \cite{betc56}. In the inviscid limit, both energy and
enstrophy are conserved. Early papers centred on the statistical consequences of these
conservation laws, and made predictions that were approximately satisfied
\cite{kraichnan67,batchelor69,kraichnan71}, but it was soon realised that another
consequence of the additional conservation law is the formation of coherent
vortex cores \cite{onsag,fornberg77,mcwilliams84}, whose presence interferes with the
regular statistical cascade \cite{mcwilliams90}. Since that realisation, the temporal evolution of
two-dimensional turbulence has mostly been described in terms of the behaviour of these
vortices \cite{mcwilliams90b,carnevale91,Benzi92}. General reviews can be found in
\cite{kraichnan80,Boff:Eck:12}

We discuss in this section how Monte-Carlo experimentation can be applied to test whether
this vortex model is the only possible one for causality in two-dimensional decaying
turbulence. The problem and the general procedure are described in
\cite{jimploff18,jimploff20}, and there are few differences between the experiments here and
in these references. The new material mostly refers to postprocessing the resulting
data, and to how conclusions can be drawn from them.
 
\begin{table}
\centering
{\begin{tabular}{llc}\hline
 Case & Perturbation to cell & Symbol\\ \hline
0&  $\omega\Rightarrow 0$ & \shortsolid\circle\\
1&  $\omega\Rightarrow -\mbox{r.m.s.}_c(\omega)$ & \shortsolid\trian\\
2&  $\omega\Rightarrow -\omega$  & \shortsolid\dtrian\\
3&  $\omega\Rightarrow \omega-\bra\omega\ket_c$, rescaled to keep r.m.s.$_c(\omega)$ constant. & \shortsolid\rtrian\\
4&  $\omega\Rightarrow -\omega+\bra\omega\ket_c$, rescaled to keep r.m.s.$_c(\omega)$ constant. & \shortsolid\ltrian\\
10&  $\uvec\Rightarrow 0$ & \dashedshort\circle\\
11&  $\uvec\Rightarrow -\mbox{r.m.s.}_c(|\uvec|)$ & \dashedshort\trian\\
12&  $\uvec\Rightarrow -\uvec$  & \dashedshort\dtrian\\
13&  $\uvec\Rightarrow \uvec-\bra\uvec\ket_c$, rescaled to keep r.m.s.$_c(|\uvec|)$ constant. & \dashedshort\rtrian\\
14&  $\uvec\Rightarrow -\uvec+\bra\uvec\ket_c$, rescaled to keep r.m.s.$_c(|\uvec|)$ constant. & \dashedshort\ltrian\\
15&  $\uvec\Rightarrow +\mbox{r.m.s.}_c(|\uvec|)$ . & \dashedshort\squar\\
\hline
\end{tabular}}
\caption{Case identification for initial perturbations. In all cases, the mean velocity and
vorticity over the full computational box are zeroed after the perturbation is applied. The signs
in the substitutions are with respect to the original cell vorticity or velocity.
Which velocity component is modified is optimised as in appendix \ref{sec:optcomp}. When
modifying $\uvec$, an extra continuity `pressure' step is applied, which may substantially
modify the perturbation. The `c' subindex labels statistics over a single cell.
}%
\label{tab:caseid}
\end{table}

The method can be summarised as follows. A number $(N_{exp})$ of experimental flow fields
(`flows' from now on) are prepared, and each of them is perturbed in various ways
(`experiments' from now on) to create initial conditions. Each experiment is run for a
prescribed time that varies between $\omega'_0 T=10$ and 30 turnovers, where $\omega'_0
=\bra \omega^2\ket ^{1/2}$ is the root-mean-square (r.m.s.) vorticity of the initial
unperturbed flow, and the time-dependent average $\bra\cdot\ket$ is taken over the full
computational box. As the evolution of the perturbed flow diverges from that of the
unperturbed initial condition, their difference is stored at several test times. The
magnitude of the evolving perturbation is defined as the norm, $\epsilon(t)$, of the
difference between the perturbed and unperturbed flow fields, evaluated over the entire
computational box, and the experiments for which this magnitude is largest are defined as
most `significant'. Four norms are used for this paper: $\|\cdot\|_2$ (quadratic) and
$\|\cdot\|_\infty$ (maximum point-wise magnitude), each of them applied to the (scalar)
vorticity or to the (vector) velocity. For each case and test time, the $n_{keep}$
perturbations with the largest and smallest deviations are classified as most or least
significant, respectively, because, in common with many complex systems, it is empirically
found that the first few most- and least-significant experiments result in fairly similar
perturbation intensities \cite{LecunEtal2015}. Because significance is defined as the effect
on the flow at some future time, the most significant experiments are also defined as being
most `causally important' for the flow evolution. In our study, the initial perturbation is
applied by dividing each flow into a regular grid of $N_c\times N_c$ square cells, each of
which is in turn modified in a number of different ways listed in table \ref{tab:caseid}. In
most cases, the results are averaged over $N_{exp}=768$ flows, on a grid
with $N_c=10$ and $n_{keep}=5$.

Simulations are performed in a doubly periodic square box of side $L=2\pi$, using a standard
spectral Fourier code dealiased by the 2/3 rule. Time advance is third-order Runge-Kutta.
The flow is defined by its velocity field $\uvec=(u,v)$ as a function of the spatial
coordinates $\xvec=(x,y)$, and time. The scalar vorticity is $\omega=\p_x v - \p_y u$, and
the rate-of-strain tensor is $s_{ij}=(\p_iu_j+\p_ju_i)/2$, where the subindices of the
partial derivatives range over $(x,y)$, and those of the velocity components over $(u,v)$.
The rate-of-strain magnitude is defined as $S^2=2 s_{ij}s_{ij}$, where repeated indices
imply summation. Time and velocity are respectively scaled with $\omega'_0$, and with $q'_0
=(u'^2 + v'^2)^{1/2}$, both measured at the unperturbed initial time, $t=0$. Unless
otherwise specified, all the cases discussed here have Fourier resolution $256^2$, with
$Re=q'_0L/\nu=2500$, where $\nu$ is the kinematic viscosity. Further details can be found in
\cite{jimploff18}.

Figure \ref{fig:grid} is a typical initial vorticity field, with the $10\times 10$ grid
overlaid. The cell outlined in red is one of the most significant ones, and the one outlined
in black is one of the less significant. Both are modified as in case 0 in table
\ref{tab:caseid} (i.e., zeroing the vorticity in the cell), and classified using the $\|
\uvec \|_2$ norm. These two cells have been chosen so that they have similar initial
perturbation intensities but fairly different intensities at the classification time, $\omega'_0
T_\sref= 4.5$. The evolution in physical space of their perturbations is shown in figure
\ref{fig:perevol}. It is clear that the difference in their temporal evolution is real, and
not a global artefact (e.g., such as a different decay rate of the turbulence intensities
in the two flows). The significant perturbation rearranges the flow in its vicinity, and
its effect eventually spreads to the full field, while the less significant one stays
localised and decays.

\begin{figure}
\centerline{%
\includegraphics[height=.32\textwidth,clip]{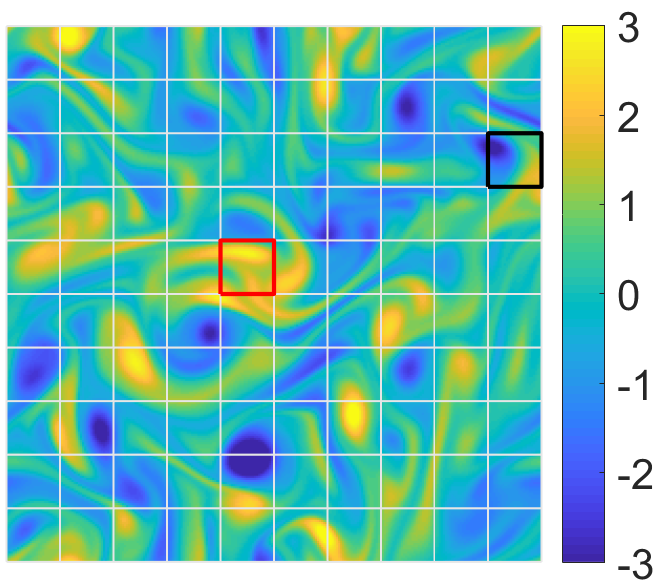}
\mylab{-.22\textwidth}{.33\textwidth}{(a)}%
\hspace*{3mm}%
\raisebox{-2mm}[0mm][0mm]{%
\includegraphics[height=.32\textwidth,clip]{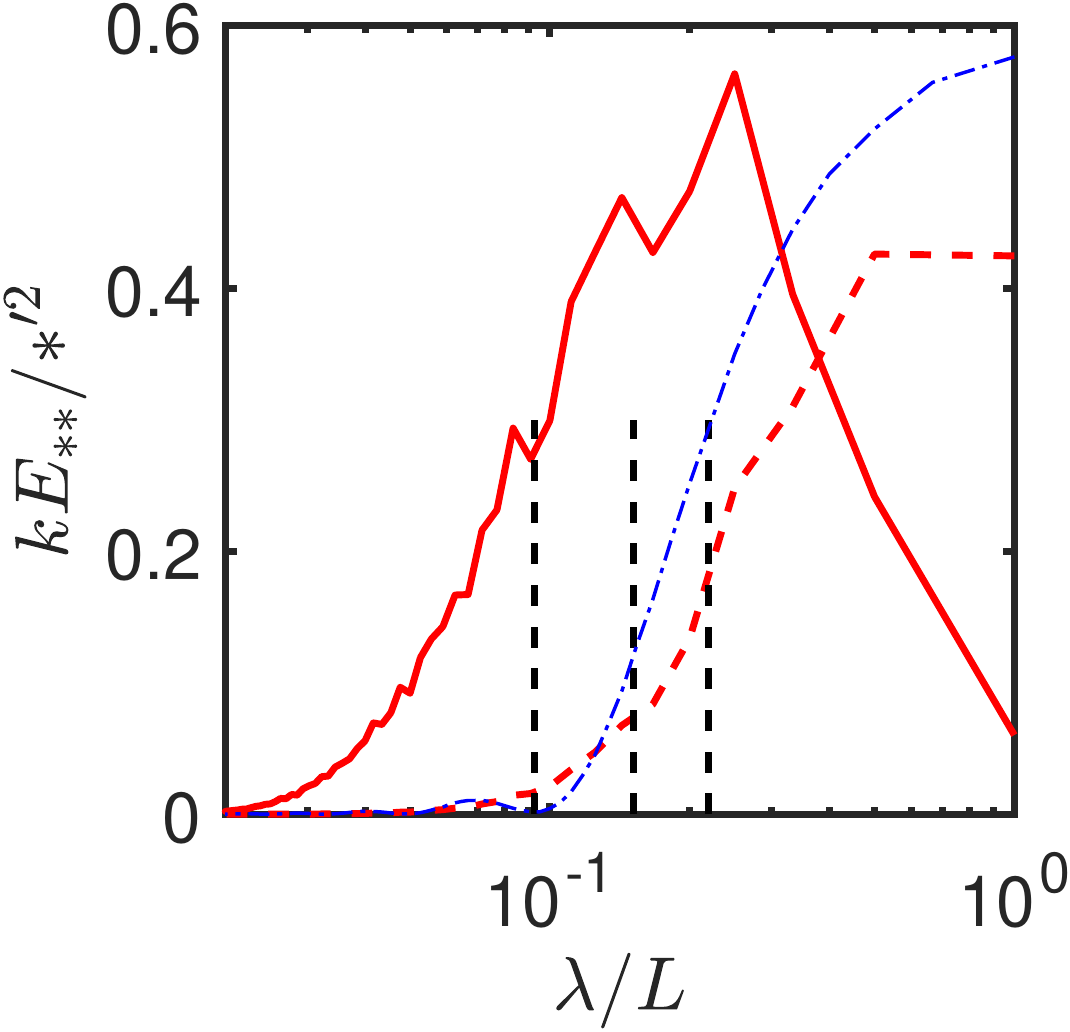}}
\mylab{-.15\textwidth}{.32\textwidth}{(b)}%
}%
\caption{(a) Initial vorticity field used for the evolutions in figure \ref{fig:perevol}.
Case 0 in table \ref{tab:caseid}, and $\| u\|_2$ norm used for classification. The cells
outlined in black (less significant) and red (more significant) have relatively similar initial
perturbation intensities, but very different later evolutions.
(b) Premultiplied spectra of the initial flow fields used for the experiments. The dashed
vertical lines are the first minimum of the transfer function of a box convolution window
corresponding to the cell size of experiments with $N_c=10$, 6 and 4, from left to right.
\solid, Enstrophy spectrum; \dashed, energy; \chndot, transfer function for $N_c=10$, scaled to fit
the plot.
} 
\la{fig:grid}
\end{figure}
%
\begin{figure}
\centerline{%
\includegraphics[height=.21\textwidth,clip]{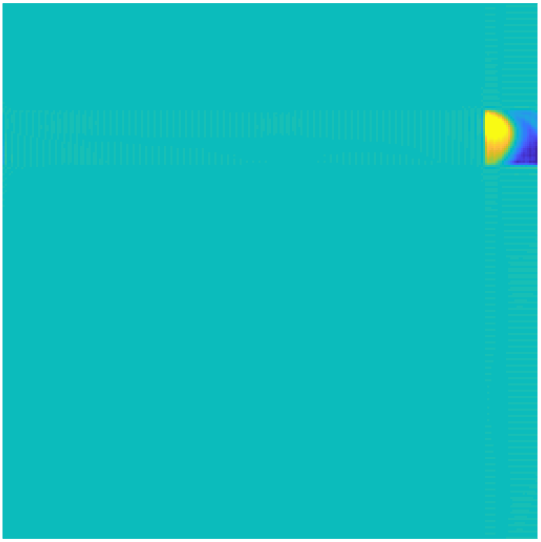}
\hspace*{1mm}%
\includegraphics[height=.21\textwidth,clip]{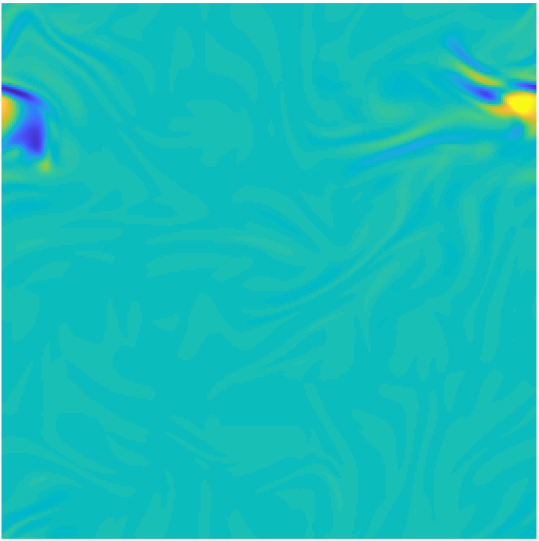}
\hspace*{1mm}%
\includegraphics[height=.21\textwidth,clip]{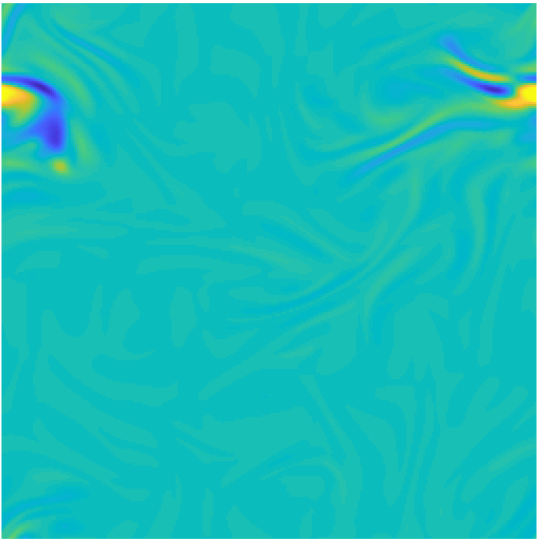}
\hspace*{1mm}%
\includegraphics[height=.21\textwidth,clip]{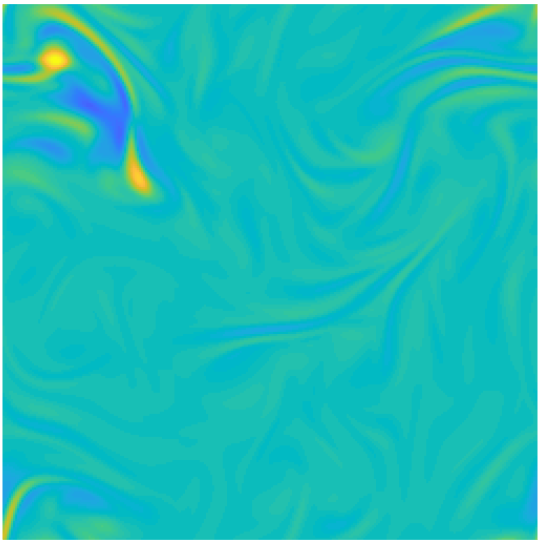}
}%
\vspace*{3mm}%
\centerline{%
\includegraphics[height=.21\textwidth,clip]{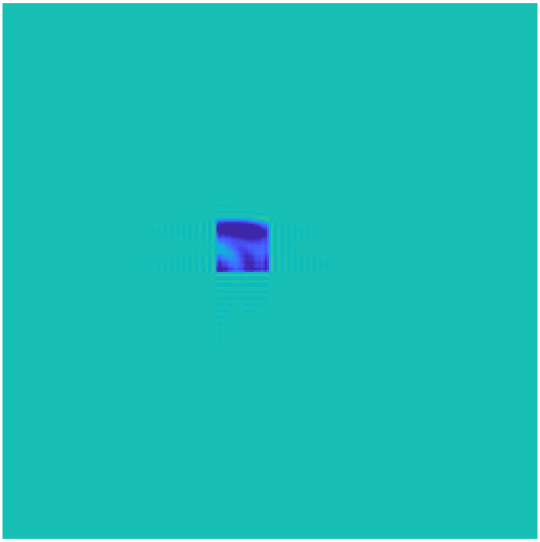}
\hspace*{1mm}%
\includegraphics[height=.21\textwidth,clip]{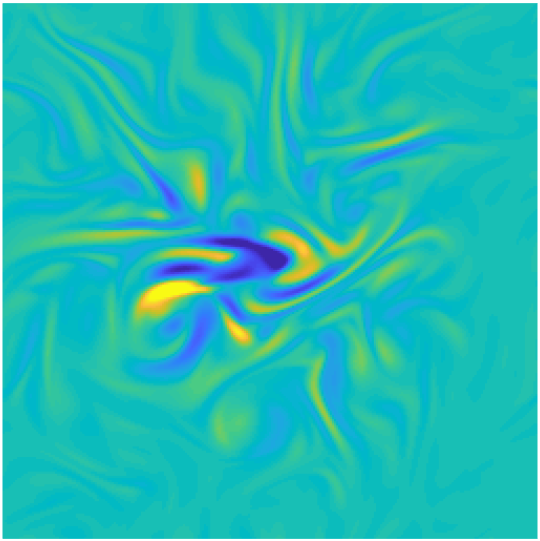}
\hspace*{1mm}%
\includegraphics[height=.21\textwidth,clip]{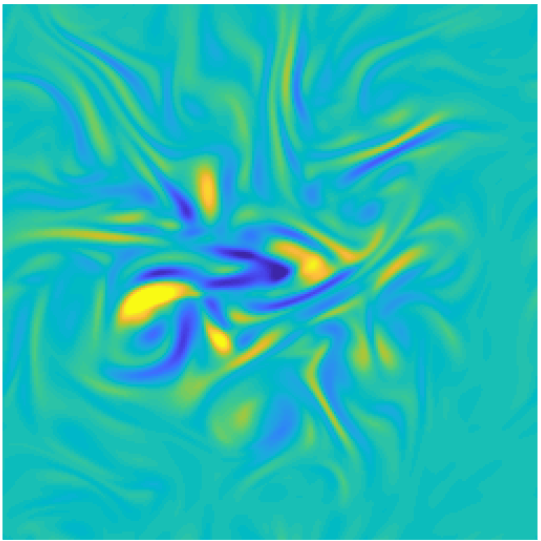}
\hspace*{1mm}%
\includegraphics[height=.21\textwidth,clip]{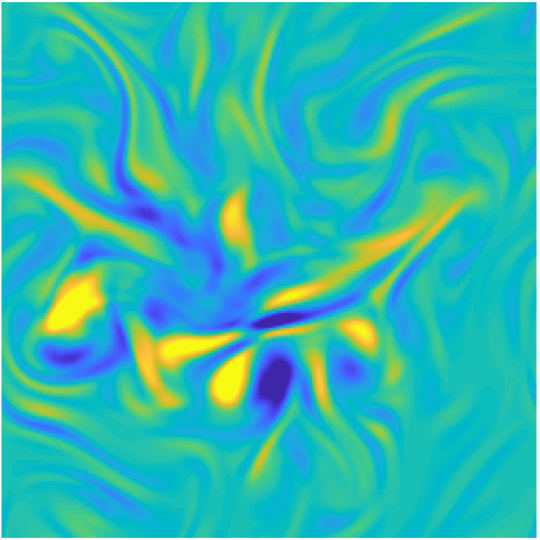}
}%
\caption{Evolution of the perturbation vorticity for the two cells marked in figure
\ref{fig:grid}. The top row is the less significant cell, marked in black in figure
\ref{fig:grid}. The bottom row is the more significant cell, marked in red. From left to
right: $\omega_0' t=0, \, 1.3,\, 2.6,\, 4.5$.
} 
\la{fig:perevol}
\end{figure}

\begin{figure}
\centerline{%
\includegraphics[width=.40\textwidth,clip]{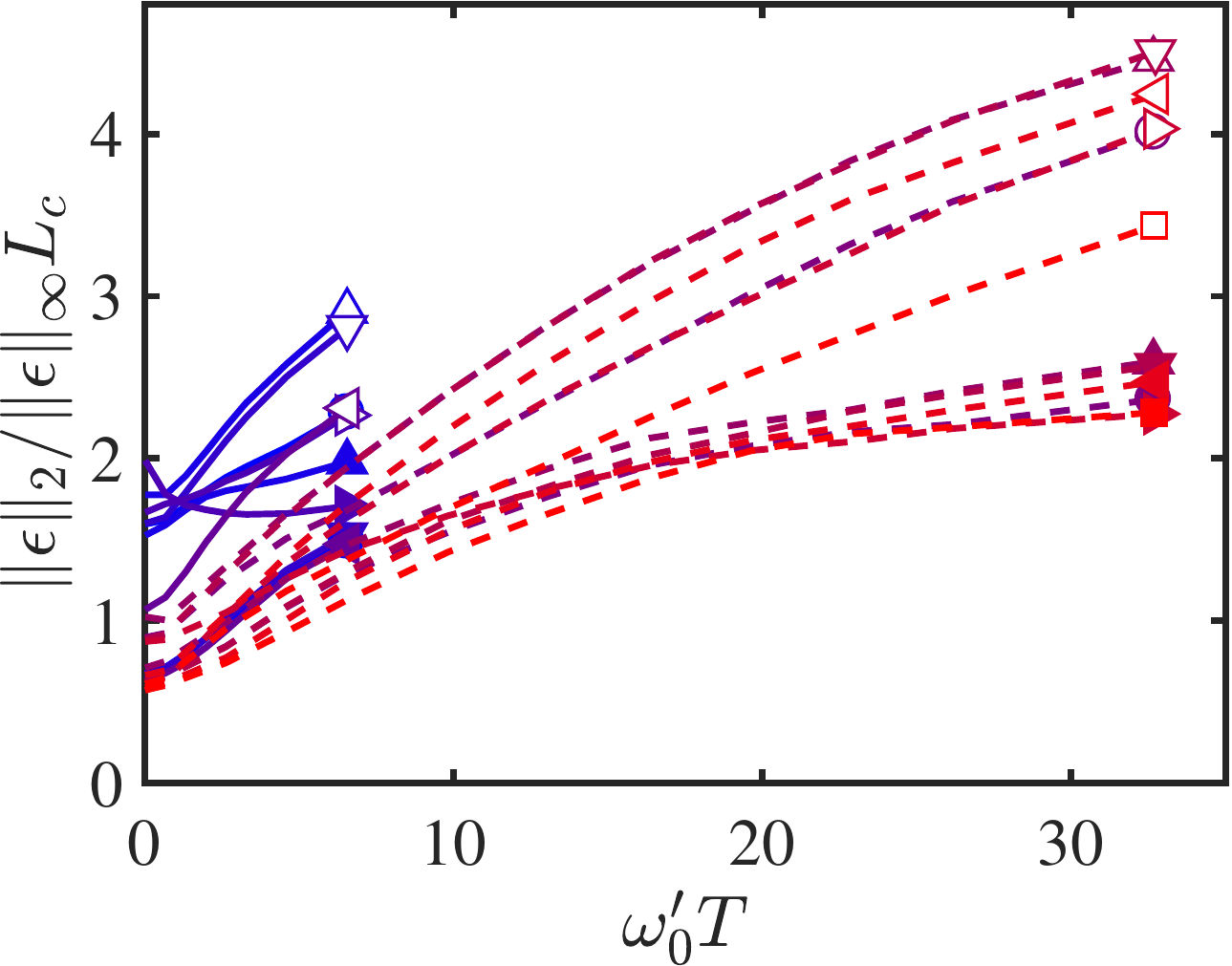}
\mylab{-.31\textwidth}{.27\textwidth}{(a)}%
\hspace*{2mm}%
\includegraphics[width=.40\textwidth,clip]{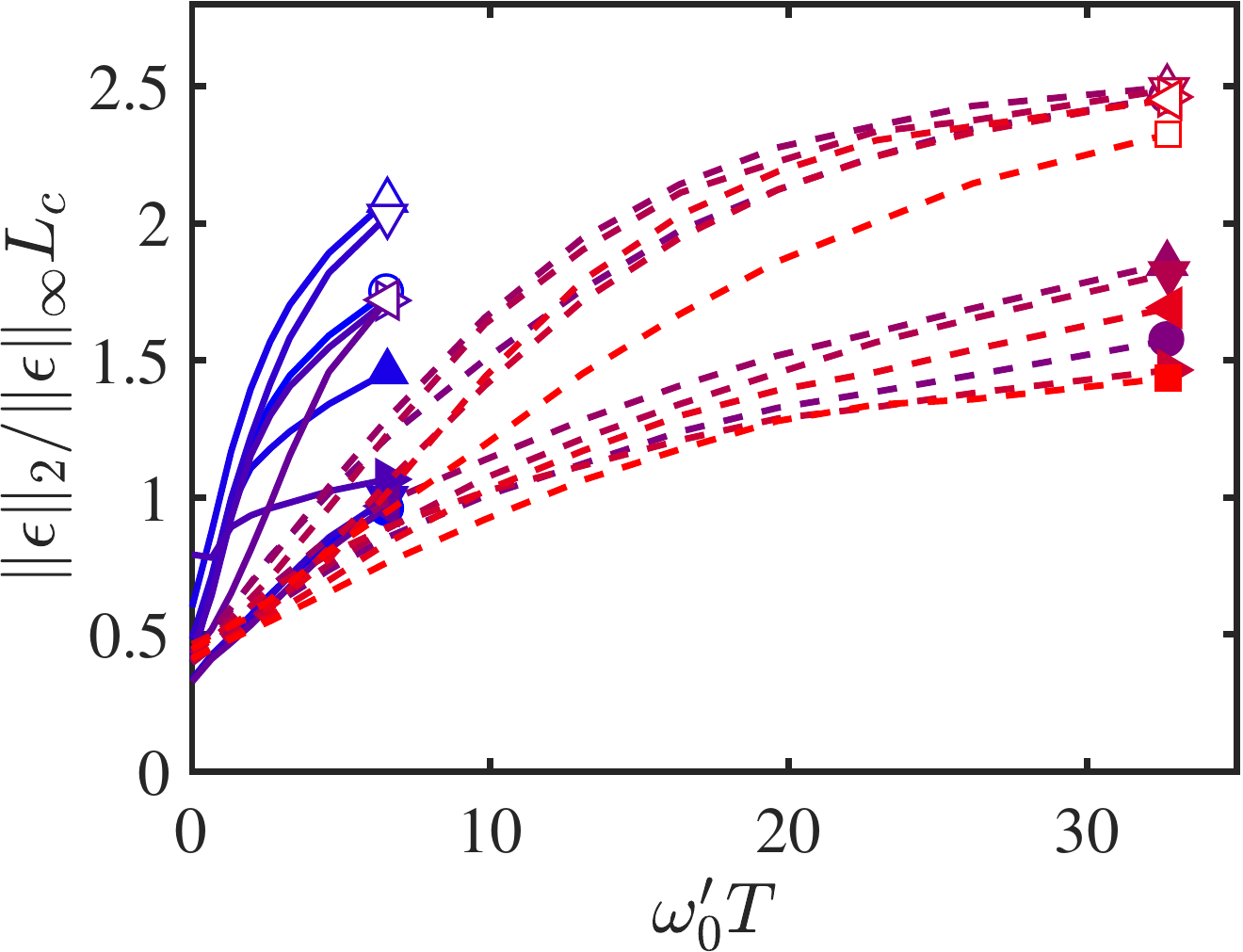}
\mylab{-.31\textwidth}{.27\textwidth}{(b)}%
}%
\caption{Geometric size of the perturbation as a function of time. Symbols as in table
\ref{tab:caseid}, but open symbols are for significant perturbations, and closed ones are
for non-significant ones. (a) Compiled using $\|\uvec\|$. (b) $\|\omega\|$. Note that the
definition of significance in this figure is applied independently each test time, so that
significant cells are not necessarily the same ones along each line (but see figure
\ref{fig:persist} below). Lines as in table \ref{tab:caseid}.
 } 
\la{fig:histsize}
\end{figure}

The spreading rate of the different perturbations is quantified in figure
\ref{fig:histsize}, which measures the geometrical size of the perturbation field as the
ratio between their integral quadratic and point-wise norms, normalised in the figure by the
size of the cells, $L_c$. Although the velocity and vorticity norms give slightly different
quantitative results, both agree that perturbations start with sizes of the order of the
cell size, and spread in times of the order of 5 -- 10 turnovers, and that significant
perturbations spread faster than non-significant ones. Recalling that $10L_c$ is the size of
the computational box, it is clear that individual perturbations have lost most of their
individuality over times, $\omega'_0 t\approx 30$, of the order of the longest experiments
in the figure.

\begin{figure}
\vspace{4mm}%
\centerline{%
\includegraphics[width=.24\textwidth,clip]{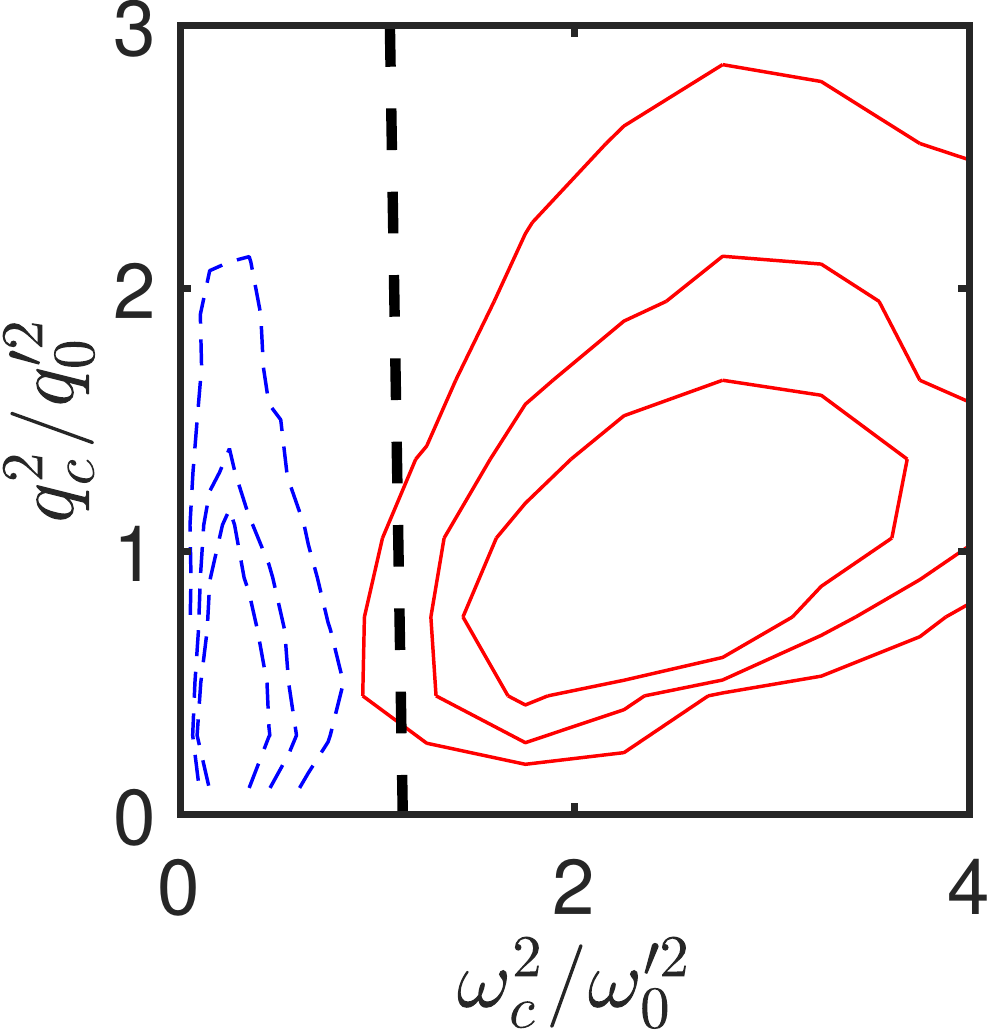}
\mylab{-0.13\textwidth}{.26\textwidth}{(a)}%
\hspace*{1mm}%
\includegraphics[width=.24\textwidth,clip]{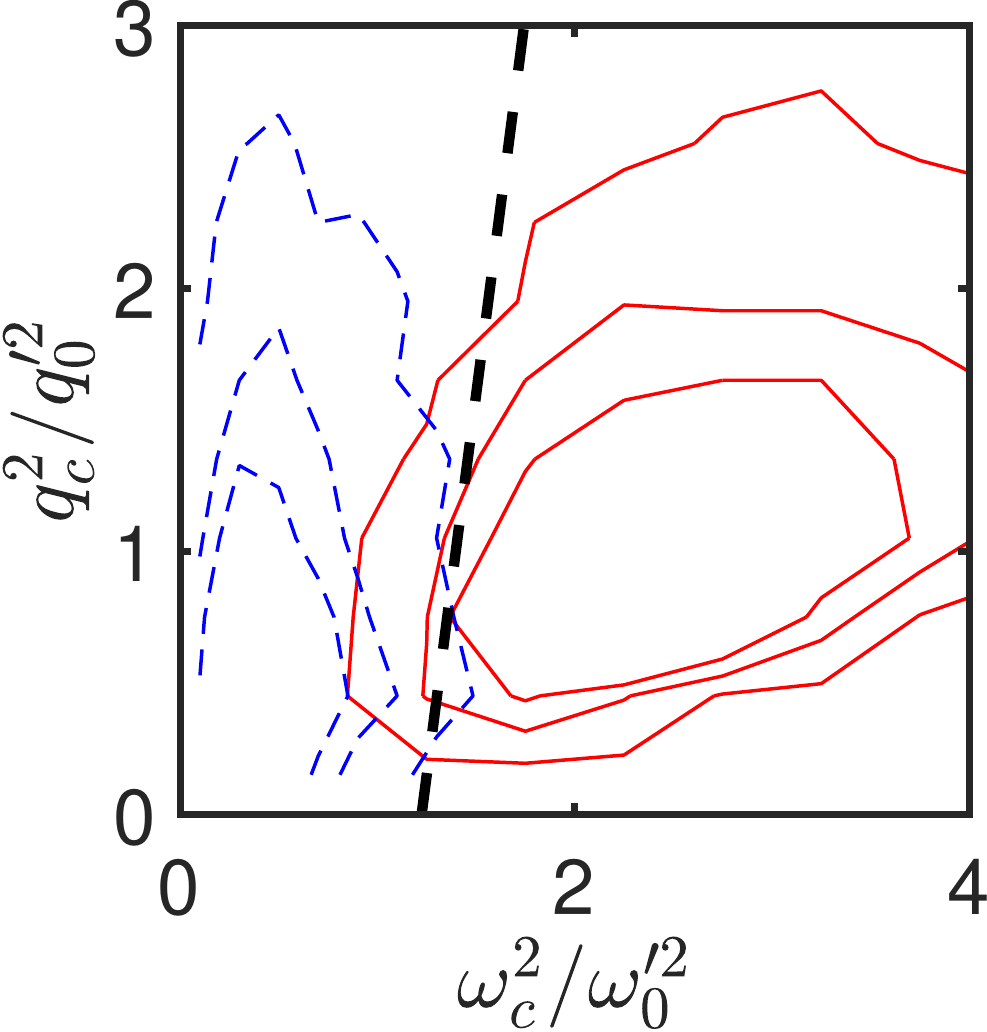}
\mylab{-0.13\textwidth}{.26\textwidth}{(b)}%
\hspace*{1mm}%
\includegraphics[width=.24\textwidth,clip]{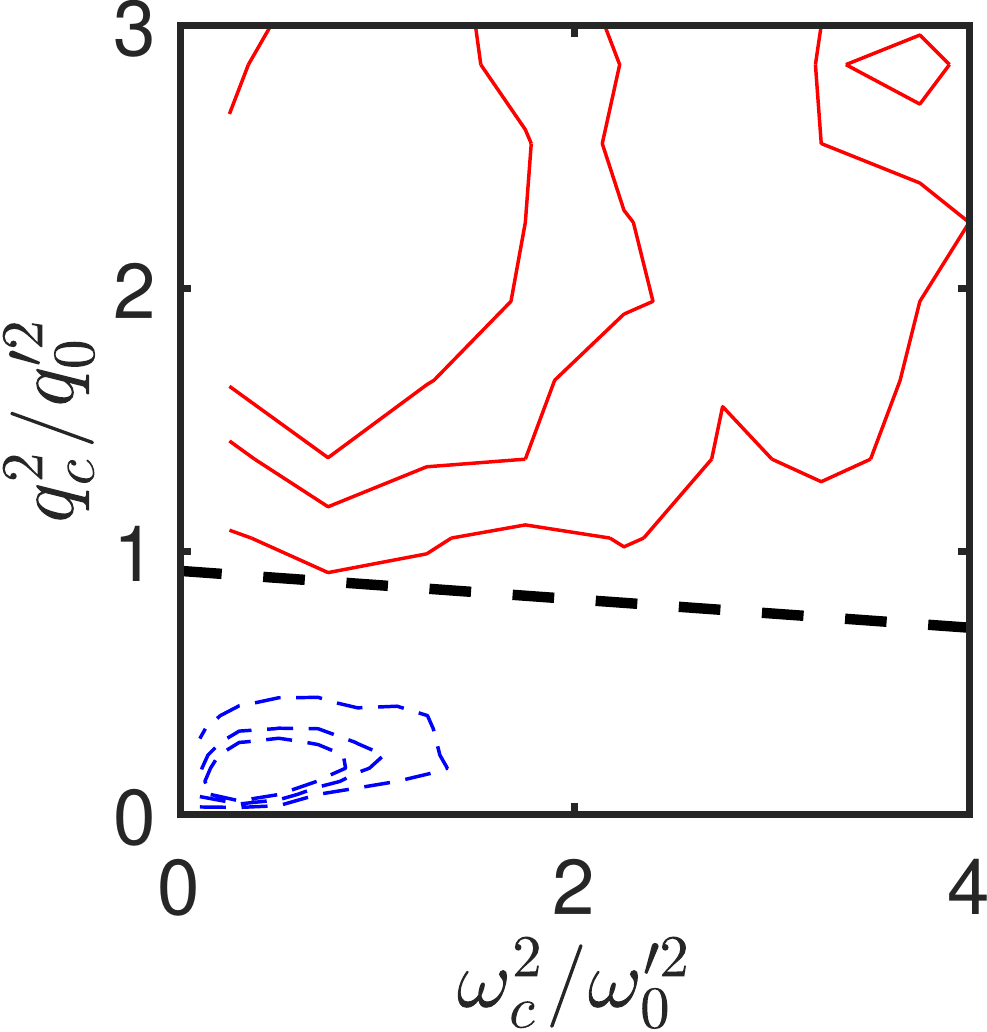}
\mylab{-0.13\textwidth}{.26\textwidth}{(c)}%
\hspace*{1mm}%
\includegraphics[width=.24\textwidth,clip]{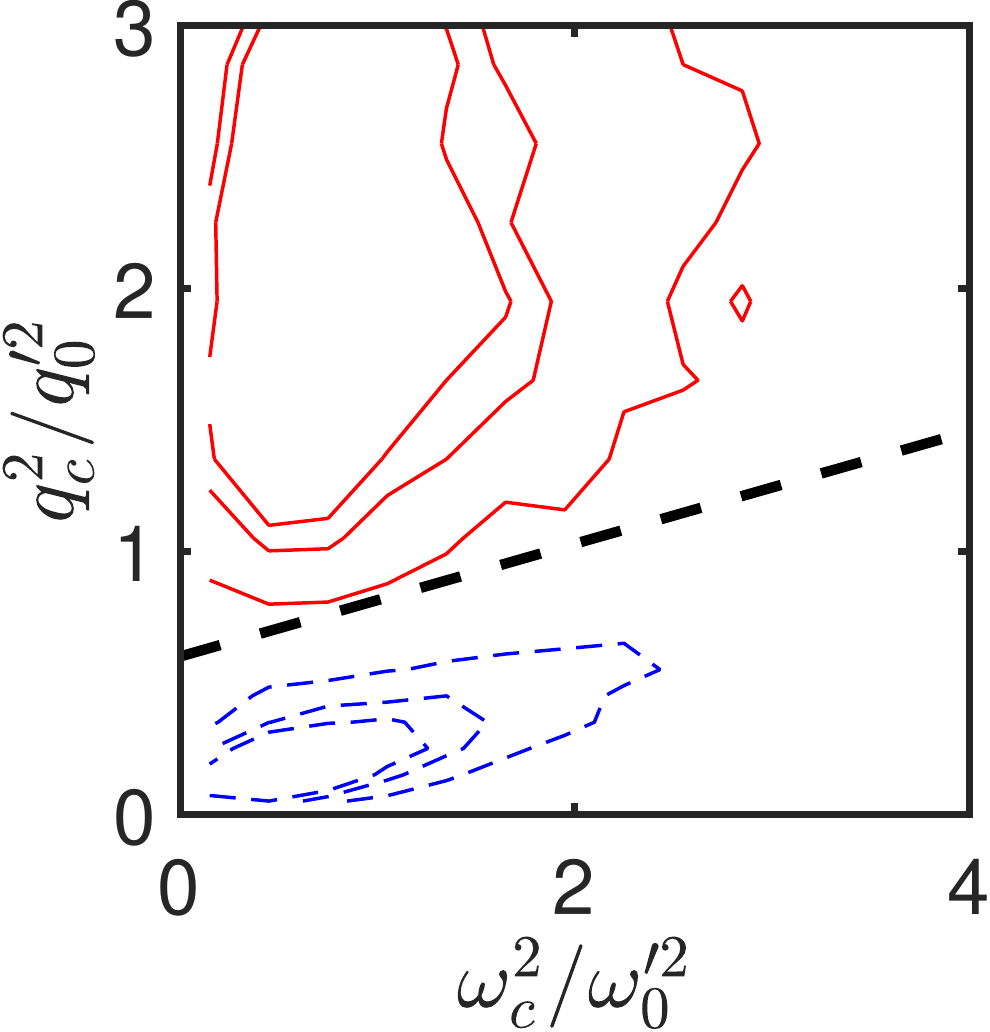}
\mylab{-0.13\textwidth}{.26\textwidth}{(d)}%
}%
\caption{Optimum classification lines for different initial perturbations, in terms of the
kinetic energy and of the enstrophy. (a) Case 0. (b) Case 3. (c) Case 10. (d) Case 13. In
all cases, 768 flows. Classification norm, $\|u\|_2$. $N_c=10$, $n_{keep}=5$. The contour
lines contain 50\%, 70\% and 90\% of the joint probability density functions (p.d.f.) of the
diagnostic variables for: \solid, most significant cases; \dashed, least significant cases.
} 
\la{fig:svm}
\end{figure}
%
\begin{figure}
\vspace*{7mm}%
\centerline{%
\includegraphics[height=.42\textwidth,clip]{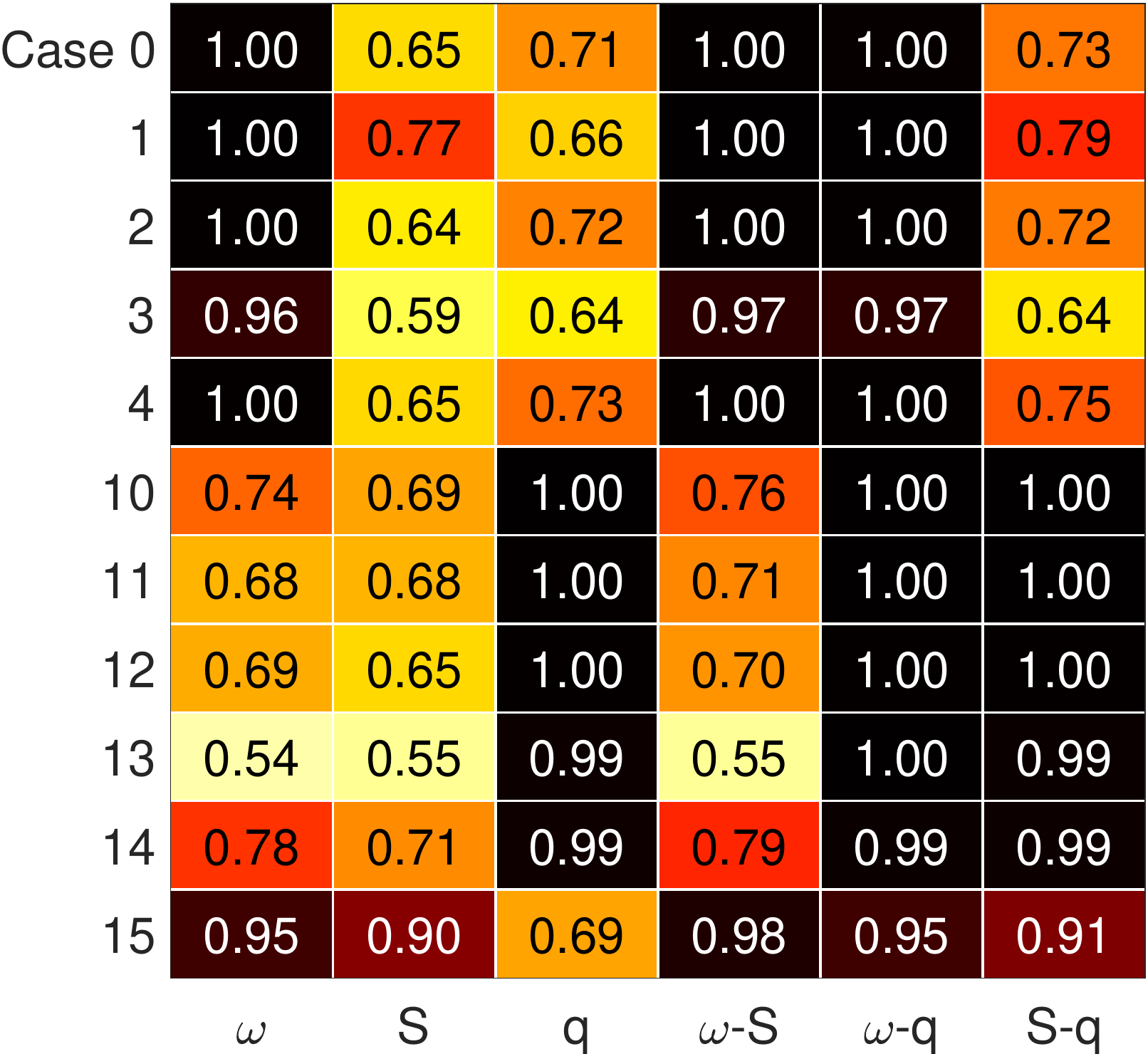}
\mylab{-0.215\textwidth}{0.44\textwidth}{(a)}%
\hspace{7mm}%
\includegraphics[height=.421\textwidth,clip]{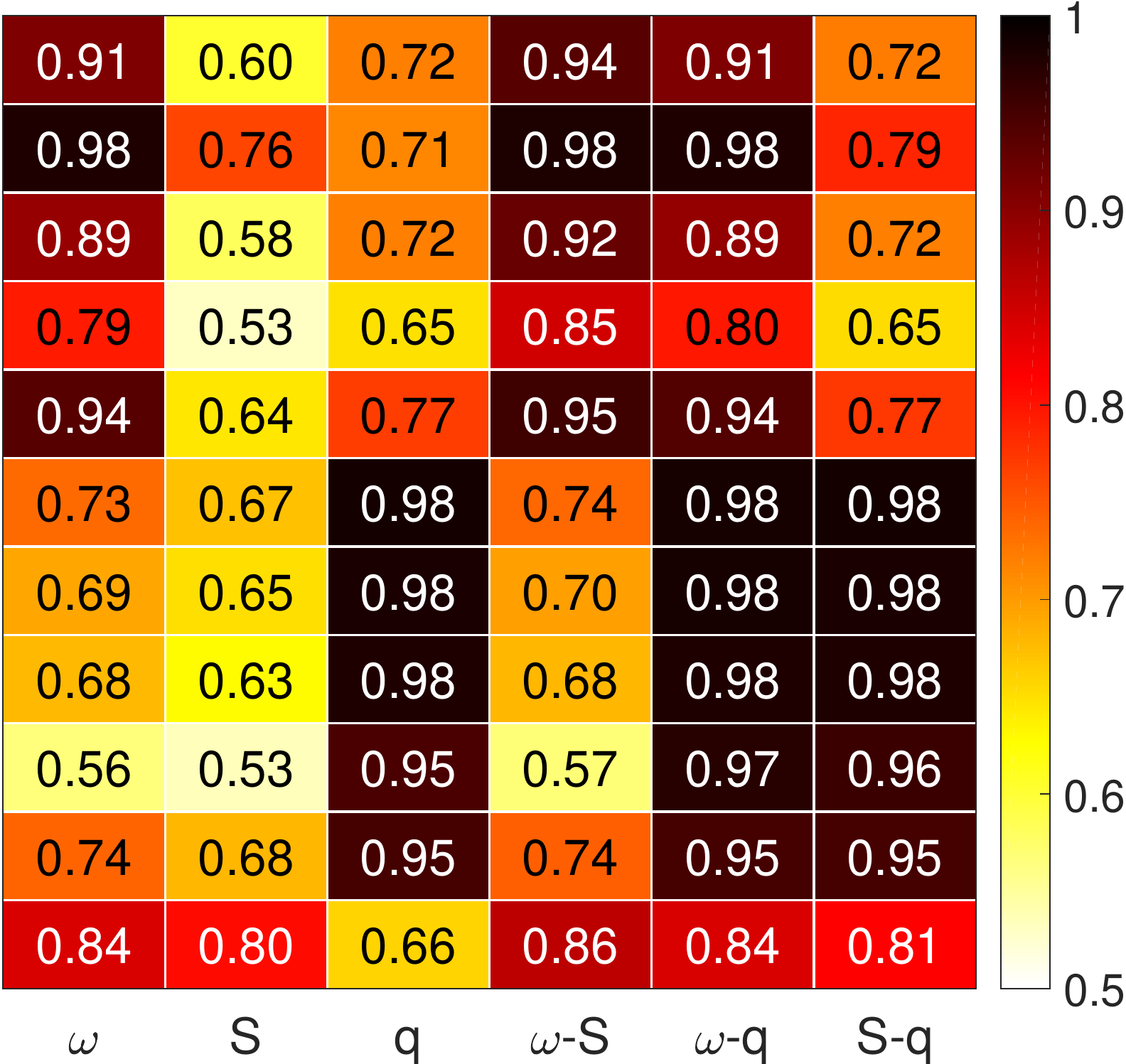}
\mylab{-0.262\textwidth}{0.44\textwidth}{(b)}%
}%
\caption{Efficiency of SVM classifiers using different combination of diagnostic variables of the
initial cells. Unit efficiency is perfect classification, and 0.5 is random guess. Rows
are the cases in table \ref{tab:caseid}, and columns are the different combinations of
diagnostic variables. Norm, $\|u\|_2$. (a) $N_c=10$. (b) $N_c=6$.
} 
\la{fig:effic}
\end{figure}

The purpose of our subsequent analysis is to determine which properties of the unperturbed
initial cells best correlate with the magnitude of the effect of modifying them. Several
factors are important, such as the modification method (table \ref{tab:caseid}), the norm
used to measure the perturbation intensity, and the time $T_\sref$ at which the
classification is done. The following analysis mostly uses the kinetic-energy norm
$\|\uvec\|_2$ as a measure of intensity, and a reference time $\omega'_0T_\sref=4.5$ for the
classification, but similar experiments were repeated using $\|\uvec\|_\infty$,
$\|\omega\|_2$ and $\|\omega\|_\infty$, with little difference in the results. The choice of
$T_\sref$ is discussed in appendix \ref{sec:tnorm}, and its effect on the results, in
\S\ref{sec:testing} and in figure \ref{fig:persist}. The properties of each cell used as
diagnostic variables are the average cell enstrophy, $\omega^2_c=\bra \omega^2\ket_c $,
defined by averaging over individual cells, the averaged kinetic energy, $q^2_c=\bra u^2+
v^2\ket_c$, and the average magnitude of the rate-of-strain tensor, $S^2_c =\bra S^2\ket_c$.
The classification scheme in references \cite{jimploff18,jimploff20} diagnosed significance
using an optimal threshold computed for each of these variables in isolation. Each cell was
classified as significant or not according to the flow behaviour at $T_\sref$,
and the resulting labelled set was used to train the threshold in such a way that
the number of misclassified events was minimised. In the present case we use a multivariable
version of the same idea (support-vector machines \cite[SVM,][]{SVM00} implemented by the
{\tt fitcsvm} Matlab routine), which determines a separating hyperplane instead of a scalar
threshold (see figure \ref{fig:svm}).

The efficiency of the classifier, defined as the fraction of correctly classified events, is
collected in the tables in figure \ref{fig:effic} for various experimental perturbations and
combinations of diagnostic variables. It is clear that some linear combination of enstrophy
and kinetic energy is always able to separate the data almost perfectly, but that $S_c$
rarely helps. The best combination of vorticity and velocity used by the classifier depends
on the initial perturbation method, as shown by the four cases in figure \ref{fig:svm}. In
general, the enstrophy is the best diagnostic variable for perturbations that manipulate the
vorticity (case 0 in figure \ref{fig:svm}.a and case 3 in figure \ref{fig:svm}.b), while the
kinetic energy is the dominant variable for the cases that manipulate the velocity (case 10
in figure \ref{fig:svm}.c and case 13 in figure \ref{fig:svm}.d). The fifth column in figure
\ref{fig:effic}(a) shows that, even in these cases, some improvement can be achieved by
including contributions from both the enstrophy and the energy, but that the effect is
marginal.

The difference among perturbation methods is best displayed by constructing for each case a
conditional `template' for the immediate neighbourhood of the significant cells in the
initial unperturbed flow. Figures \ref{fig:conditional}(a,b) include templates for cases 0
and 10, built from the $3\times3$-cell neighbourhood of the most significant cell in each
experiment. To take into account the reflection and rotational symmetries of the equations
of motion, the template is computed by averaging these flow patches after rotating and
reflecting them so that they mutually agree as much as possible. To compensate for the
effect of the magnitude of the templates, their intensity is scaled to match the global intensity of
the flow before comparing them to individual neighbourhoods.

Figure \ref{fig:conditional}(a), which represents the conditional structure in the vicinity
of cells that are most sensitive to zeroing their vorticity, is an isolated vortex. This
would appear to support the classical view that two-dimensional turbulence is 
controlled by the interactions among individual vortices \cite{mcwilliams90b,carnevale91}.
But the template in figure \ref{fig:conditional}(b), which corresponds to cells that have
been perturbed by zeroing their velocity, and which are thus best diagnosed by the magnitude of
their kinetic energy, is a vortex dipole. This is a less expected result, but a reasonable
one, because the velocity field of a dipole contains a local jet, and it makes sense that blocking
it has a strong effect. In general, the templates for the most significant structures in
experiments that manipulate the vorticity, are isolated vortices, while the manipulation of the
velocity results in dipoles.

\begin{figure}
\vspace{5mm}%
\centerline{%
\includegraphics[height=.25\textwidth,clip]{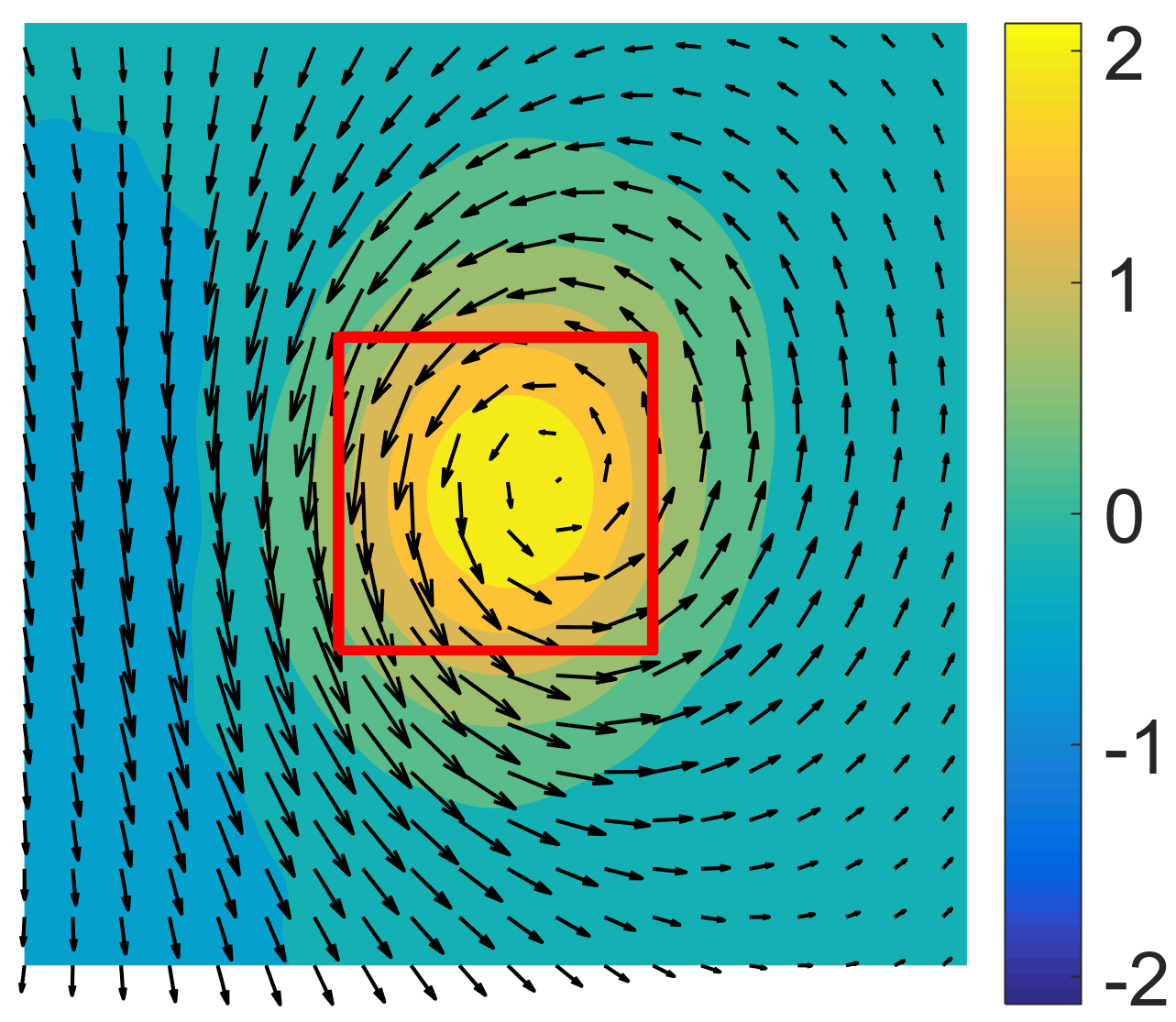}
\mylab{-.33\textwidth}{.20\textwidth}{(a)}%
\hspace*{9mm}%
\includegraphics[height=.25\textwidth,clip]{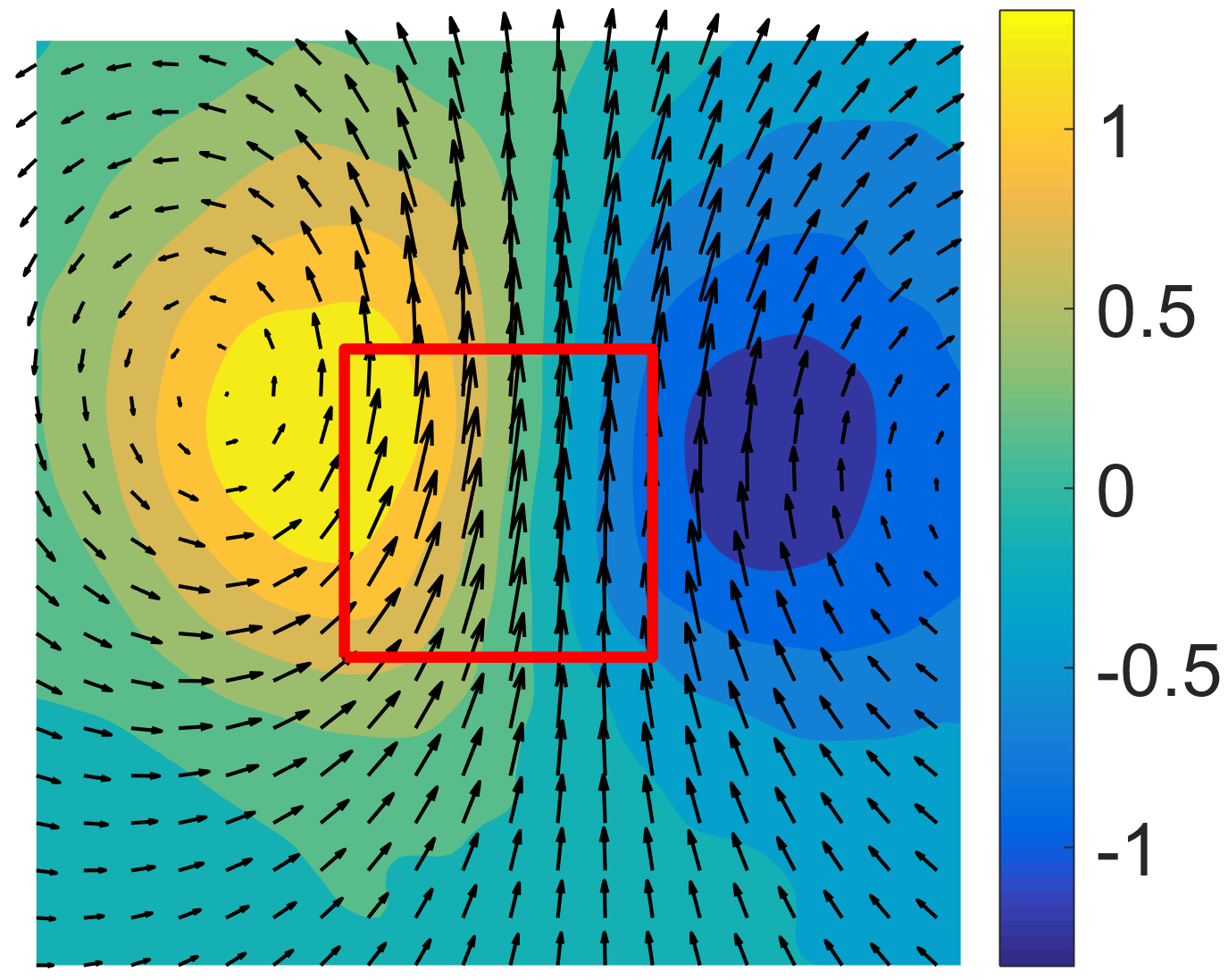}
\mylab{-.35\textwidth}{.20\textwidth}{(b)}%
}%
\vspace{3mm}%
\centerline{%
\includegraphics[height=.25\textwidth,clip]{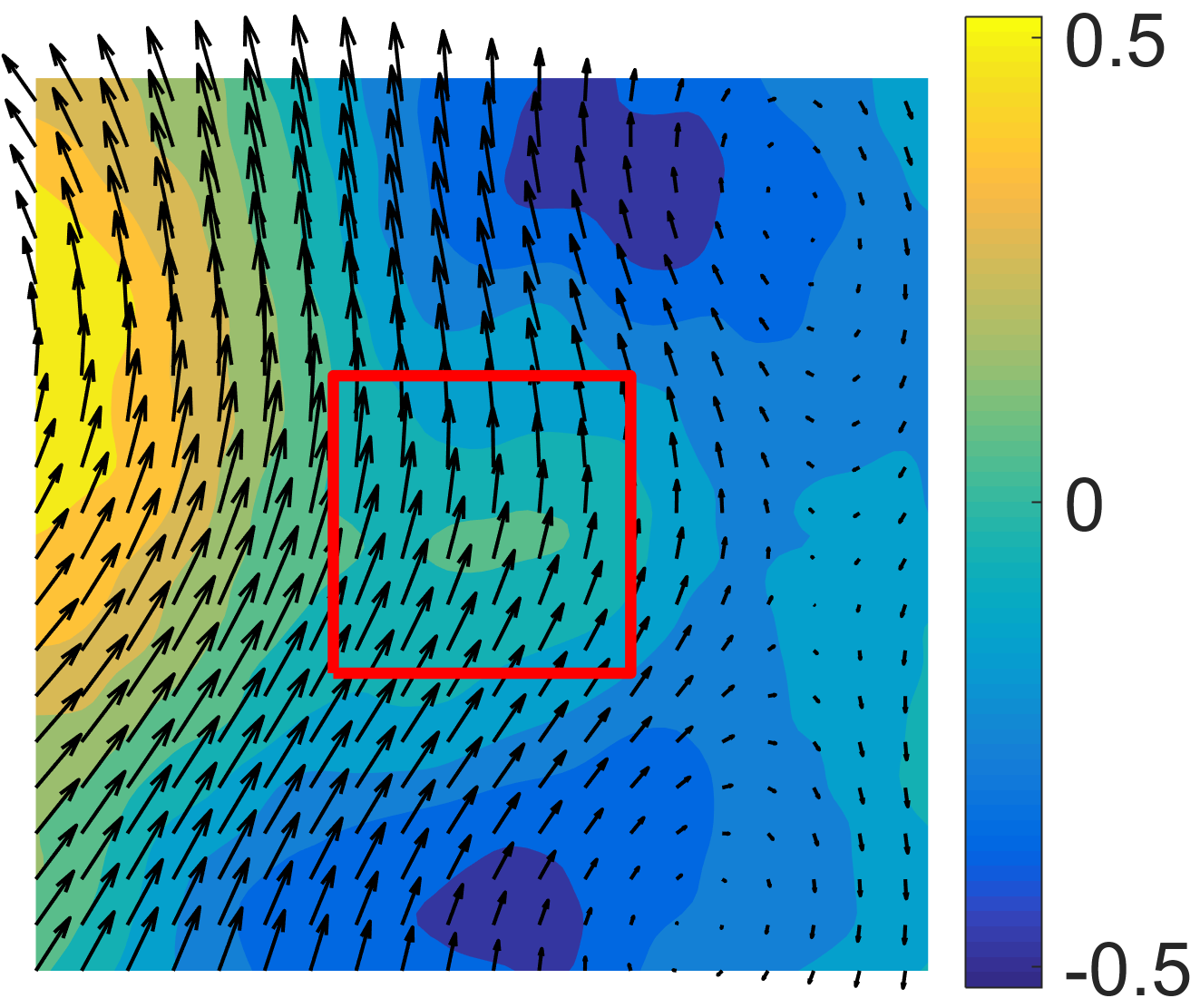}
\mylab{-.35\textwidth}{.20\textwidth}{(c)}%
\hspace*{8mm}%
\includegraphics[height=.25\textwidth,clip]{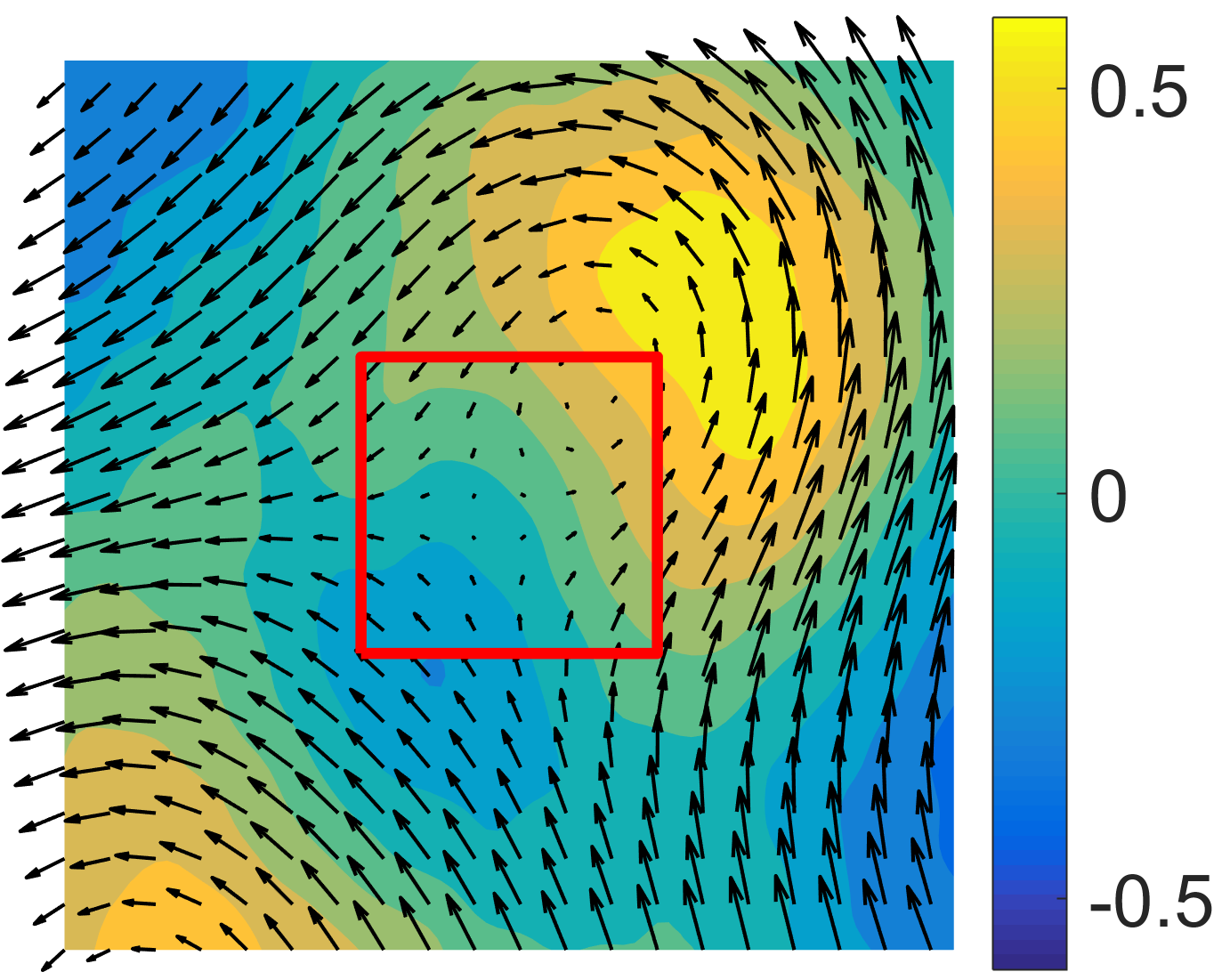}
\mylab{-.34\textwidth}{.20\textwidth}{(d)}%
}%
\caption{Conditional vorticity and velocity distributions, normalised with their
unconditional r.m.s., in the neighbourhood of the most (top row) and least (bottom)
significant cells. The test cell is outlined in red and, because of the invariances of the
problem, the orientation and the vorticity sign are immaterial. (a,c) Case 0 in table
\ref{tab:caseid}. (b,d) Case 10. Compiled from 768 flows, and classified using $\| \uvec
\|_2$.
 } 
\la{fig:conditional}
\end{figure}

It was mentioned in \cite{jimploff20} that case 15 in table \ref{tab:caseid} is an exception
to this rule, because figure \ref{fig:effic} shows that, even if this experiment involves
manipulating the velocity, its most diagnostic variable is the enstrophy. A little
reflection shows the reason. What this experiment does is to substitute the velocity field
in the cell by a uniform velocity equal to the average over the original cell. At such, it
removes the velocity fluctuations without modifying the mean, and is closer to a vorticity
manipulation than to a velocity one. Correspondingly, the conditional template for this case
is an isolated vortex. 

Figures \ref{fig:conditional}(c,d) show templates conditioned to neighbourhoods of the least
significant cells. They are less clear and weaker than the significant ones, and they vary
more among different experimental conditions (see \S\ref{sec:verify}).

The classification efficiency and the resulting templates depend on the size of the
experimental cells. In general, the efficiency degrades for larger cells, as shown in figure
\ref{fig:effic} for $N_c=10$ and $N_c=6$, and continues to do so for $N_c=4$ (not shown). In
the case of enstrophy, this evolution is consistent with the spectrum in figure
\ref{fig:grid}(b), which shows that the cell size when $N_c=10$ is of the order of the
vortex size, but it is a little surprising that fairly small cells work so well for the
kinetic energy, whose spectrum peaks at the scale of the box, suggesting that, at least at
this low Reynolds number, the causality of the kinetic energy remains concentrated at the
scale of individual vortex pairs (see \S\ref{sec:physics} for a more complete discussion).
In fact, the effectiveness of the enstrophy and of the energy behave differently with the
cell size. While the mean effectiveness of $\omega_c^2$ as a diagnostic variable in cases
0--4 decays from 0.99 to 0.74 as $N_c=10 \to 4$, that of $q_c^2$ in cases 110--114 only
decays from 0.99 to 0.88 in the same range. The templates obtained from coarser experimental
grids are similar to those for $N_c=10$, but become progressively less well defined as $N_c$
decreases.

\subsection{Verification}\la{sec:verify}

Any result based on the exploitation of large data sets is statistical and, as such, needs
to be verified and validated. Verification refers to whether the experimental procedure can
be trusted, which mostly has to do with avoiding statistical artefacts. It is discussed in
this section. Validation assesses whether the results are relevant to the problem at hand,
and is deferred to \S\ref{sec:testing}.

One effect of conditioning can be seen from the difference between the colour bars in figure
\ref{fig:conditional} and those of the full flow in figure \ref{fig:grid}(a). While the
instantaneous vorticity in the latter spans the range $\omega/\omega'_0\in \pm 3$, the
significant templates in figures \ref{fig:conditional}(a,b) only span
$\omega/\omega'_0\in \pm$(1--2). The r.m.s. intensity, $\| \uvec_{T} \|_2/q'_0$, of
the different templates is given in the first column of table \ref{tab:temperr}, where the
norm is computed over the $3\times 3$ domain of each template in two representative cases (a
vortex template in case 0, and a dipole in case 10). The two numbers in each line of this
table apply to the templates of the most and least significant cells in each experiment,
respectively corresponding to figures \ref{fig:conditional}(a,b) and
\ref{fig:conditional}(c,d). Results are essentially the same for other cell modification
methods.

The third line in table \ref{tab:temperr} refers to templates computed from randomly chosen
cells, and is similar to the least significant cases in the lines
above. It is interesting that the intensities in this line are as high as they are. The
naive expectation would be that averaging 768 random cells of individual intensity $q'$
would result in a standard deviation $q'/\sqrt{768} \approx 0.036 q'$, but the bottom line
of table \ref{tab:temperr} is ten times higher than this. Clearly, the process of optimally
aligning the individual flows while computing the templates creates a template
even when there is none. It follows from table \ref{tab:temperr} that the templates created
from the least significant cells are only as good as random, while those for the most
significant ones are almost as strong as the unconditional flow.

\begin{table}
\centering
{\begin{tabular}{lccc} \hline
Template  &\parbox{3cm}{\centering  Template intensity\\$\|\uvec_T\|_2/q'_0$} 
&\parbox{3cm}{\centering Convergence error\\(128 -- 768)}
&\parbox{3cm}{\centering  Approximation error\\(Flow -- Template)} 
\\ \hline
Vortex (case 0)& 0.86 -- 0.54 &  0.23 -- 0.55  & 0.71 -- 0.98\\
Dipole (case 10)& 0.95 -- 0.43 & 0.21 -- 0.42   & 0.66 -- 1.13  \\
Random&  0.56 &  0.37 & 0.97\\
\hline
\end{tabular}}
\caption{Measures of approximation error and statistical convergence for the templates in figure
\ref{fig:conditional}. 
The template intensity is $\| \uvec_{T} \|_2$, defined over its $3\times 3$
box.
The convergence error is defined as the difference between templates
computed from 768 and 128 experiments,
$\|\uvec_{T(128)}-\uvec_{T(768)}\|_2/\|\uvec_{T(768)}\|_2$.
The approximation error is the  difference between the templates and the flow,
$\|\uvec-\uvec_{T}\|_2 / \|\uvec\|_2$. 
Except for the second column, data are averaged over 768 experiments.
}%
\label{tab:temperr}
\end{table}

The second column of table \ref{tab:temperr} confirms this conclusion. It shows the 
difference between templates computed from relatively few experiments (128) and those
constructed from the full set of experiments (768). For the reasons explained when
discussing the template intensity, it is difficult to predict which should be the baseline
difference between the results of unrelated experiments, but it is clear from the table that
the significant templates -- the first of the two numbers in the first two lines -- are
significantly more converged than the least significant ones, strongly suggesting that the
former are physically significant properties, while the latter are not.

The last column in table \ref{tab:temperr} is the relative approximation error between the
flow and the template. As above, the first number in the first two lines is for the most
significant templates, and the second one is for the least significant ones. The last line
is for random cells, and its approximation error is consistent with the null template that
would be obtained from adding independent cells, although we saw above that the random
templates are weak, but not null. The approximation errors for the significant cells are
lower than random, but still substantial. They show that, although templates are a
reasonable approximation to the structure of the flow in the neighbourhood of significant
cells, we should not expect to find many `pure' circular vortices or dipoles in the flow
(see figure \ref{fig:grid}.a).
 
 \subsection{Validation}\la{sec:testing}

Although it should be clear by now that the templates in figure \ref{fig:conditional}(a,b)
contain structural information about the significant flow regions defined above, and that
they can be statistically trusted, it remains to be shown that they represent flow
properties rather than experimental artefacts. Moreover, although the previous analysis
shows that the most significant perturbations are preferentially those which are applied to
the neighbourhood of vortices and dipoles, it is unclear whether the converse is also true,
and flow regions that can be approximated as vortices and dipoles are preferentially
significant.

\begin{figure}
\centerline{%
\includegraphics[width=.40\textwidth,clip]{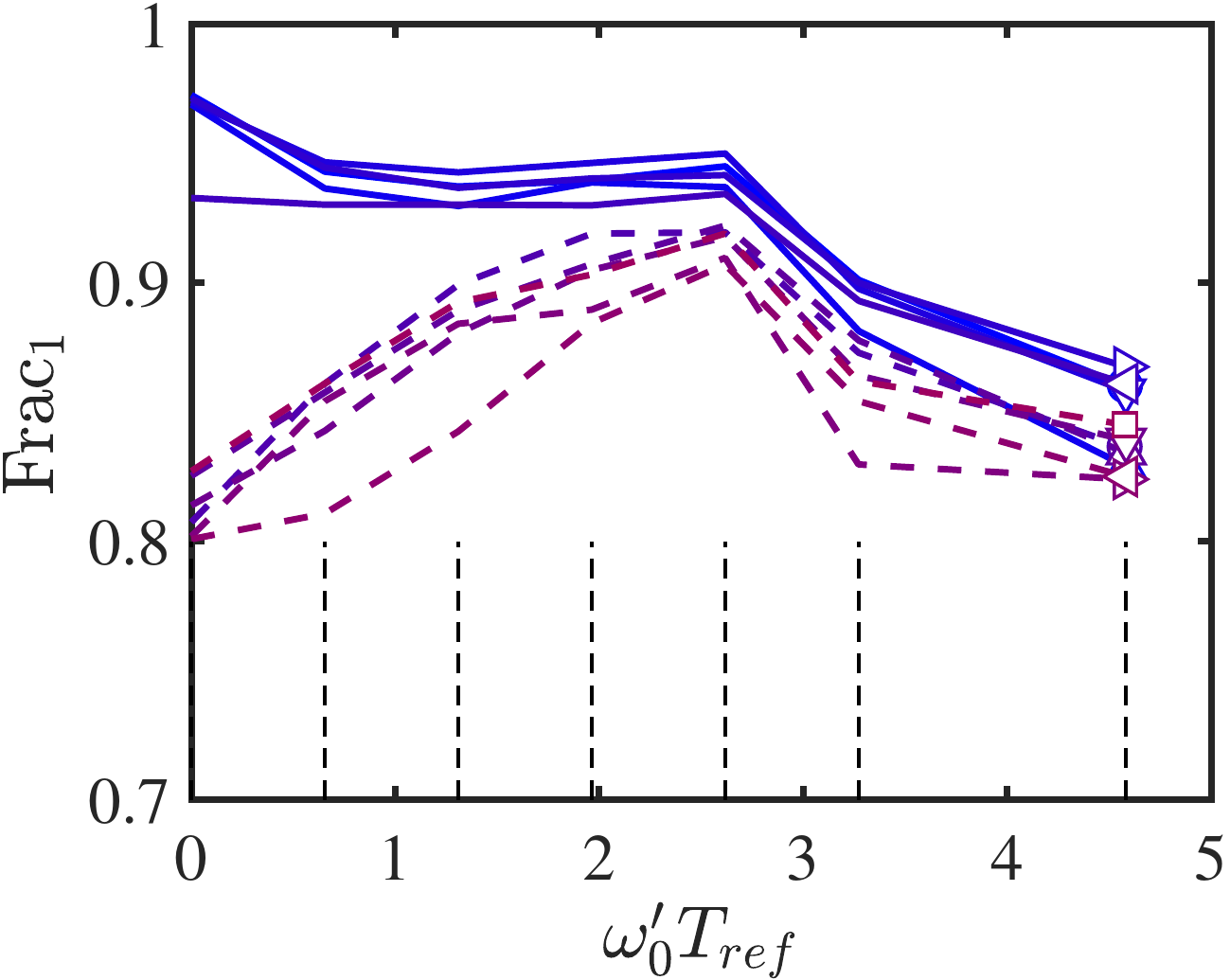}
\mylab{-.07\textwidth}{.27\textwidth}{(a)}%
\hspace*{2mm}%
\includegraphics[width=.40\textwidth,clip]{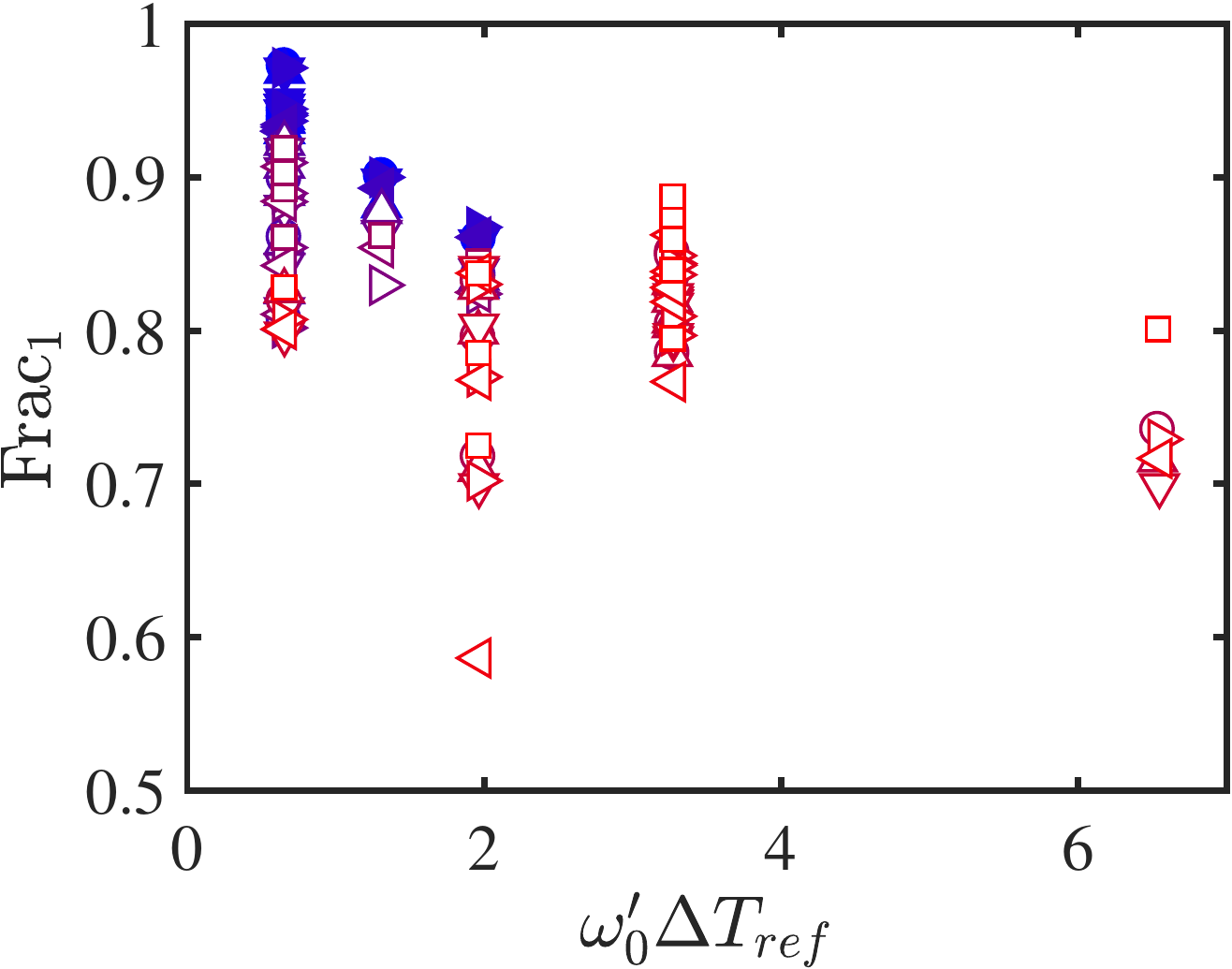}
\mylab{-.07\textwidth}{.27\textwidth}{(b)}%
}%
\caption{(a) Fraction of cells in the initial flow classified as significant when measured
at one test time, and still classified as such at the next test time,
as a function of time. Test times are marked by the dashed vertical lines.
(b) As in (a), but as a function of the time difference between consecutive test times. Symbols as in
table \ref{tab:caseid}, but closed symbols in (b) are vorticity perturbations (cases 0--4)
and open symbols are velocity perturbations (cases 10--15). In both figures, the
significance norm is $\|\uvec\|_2$. Lines and symbols as in table \ref{tab:caseid}.
 } 
\la{fig:persist}
\end{figure}

Consider first the sensitivity of the results to experimental conditions. We have already
seen that different experiments produce different templates, but this is not particularly
worrisome, and should be considered a desirable property of the Monte-Carlo exploration of
possibilities. More disturbing is the dependence on the testing time. The experiments above
produce a classification of flow features at time $T=0$, based on the intensity of their
perturbations at a later time, $T_\sref$. The choice of $T_\sref$ is discussed in
appendix \ref{sec:tnorm}, where it is shown that there is a characteristic period, of the
order of $\omega'_0 T=5$, over which causality develops and is preserved. But, even within
this constraint, the classification depends on $T_\sref$, and it is important to
determine whether cells classified as significant for a given $T_\sref$, remain significant
when classified at a different time.

This persistence is checked in figure \ref{fig:persist}(a), which shows the fraction of
cells classified as significant at one of the test times stored by the experiment, which are
still classified as such at the next test time. The fraction is high (random sets would only
persist 0.25\% of the time), and figure \ref{fig:persist}(b) shows that the decay is mostly a
function of the interval among consecutive test times. In agreement with the results in
appendix \ref{sec:tnorm}, figure \ref{fig:persist} suggests an effective persistence time of
the order of a few turnovers, and indicates that this is the correct time interval over
which to study causality in this flow. The results are approximately independent of the norm
used for the classification, although figure \ref{fig:persist} shows that vorticity norms
are slightly less persistent than velocity ones, and it can be shown that quadratic norms
are slightly more persistent than point-wise ones.

\begin{figure}
\centerline{%
\includegraphics[width=.43\textwidth,clip]{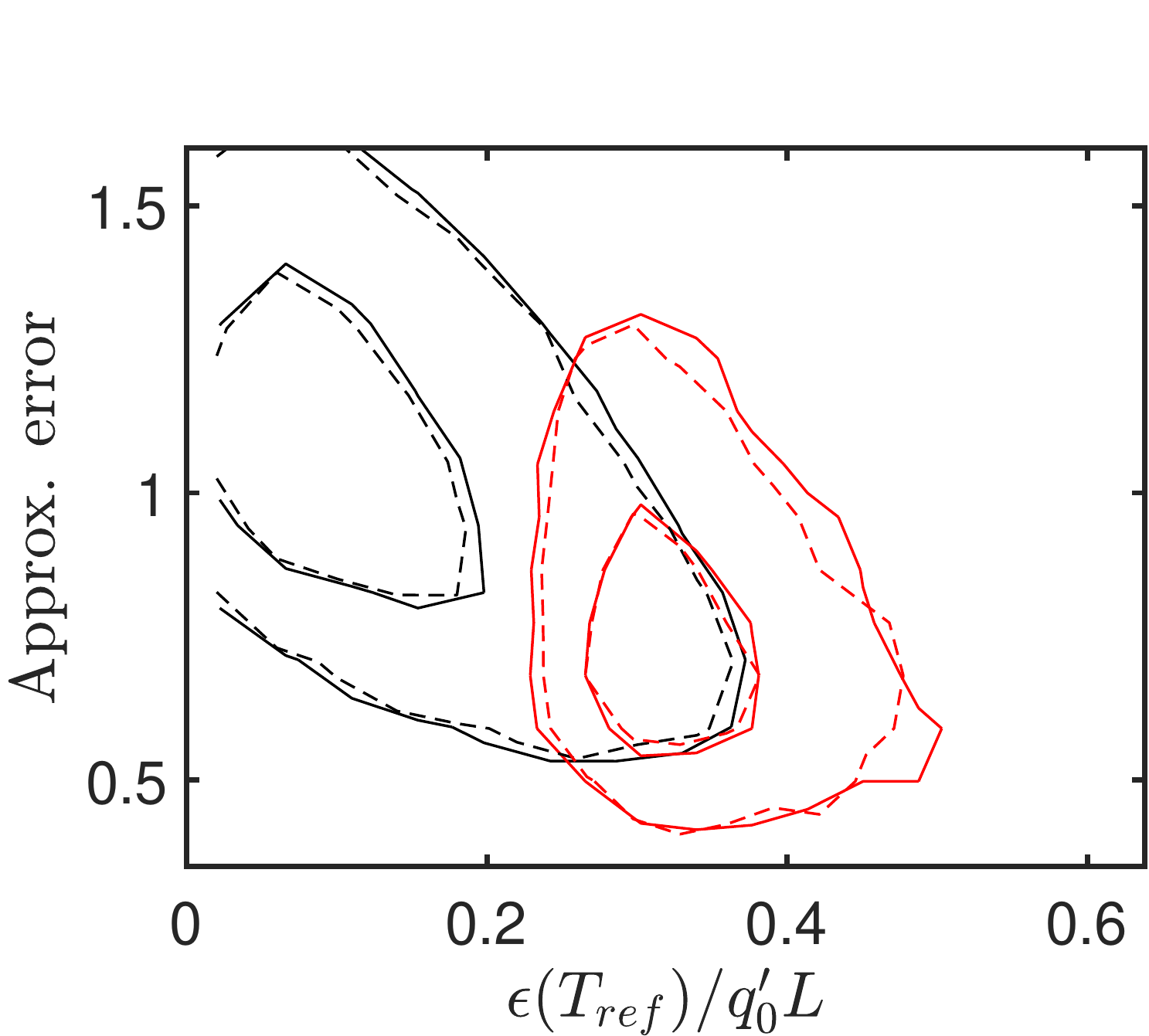}%
\mylab{-.07\textwidth}{.28\textwidth}{(a)}%
\hspace*{5mm}%
\includegraphics[width=.43\textwidth,clip]{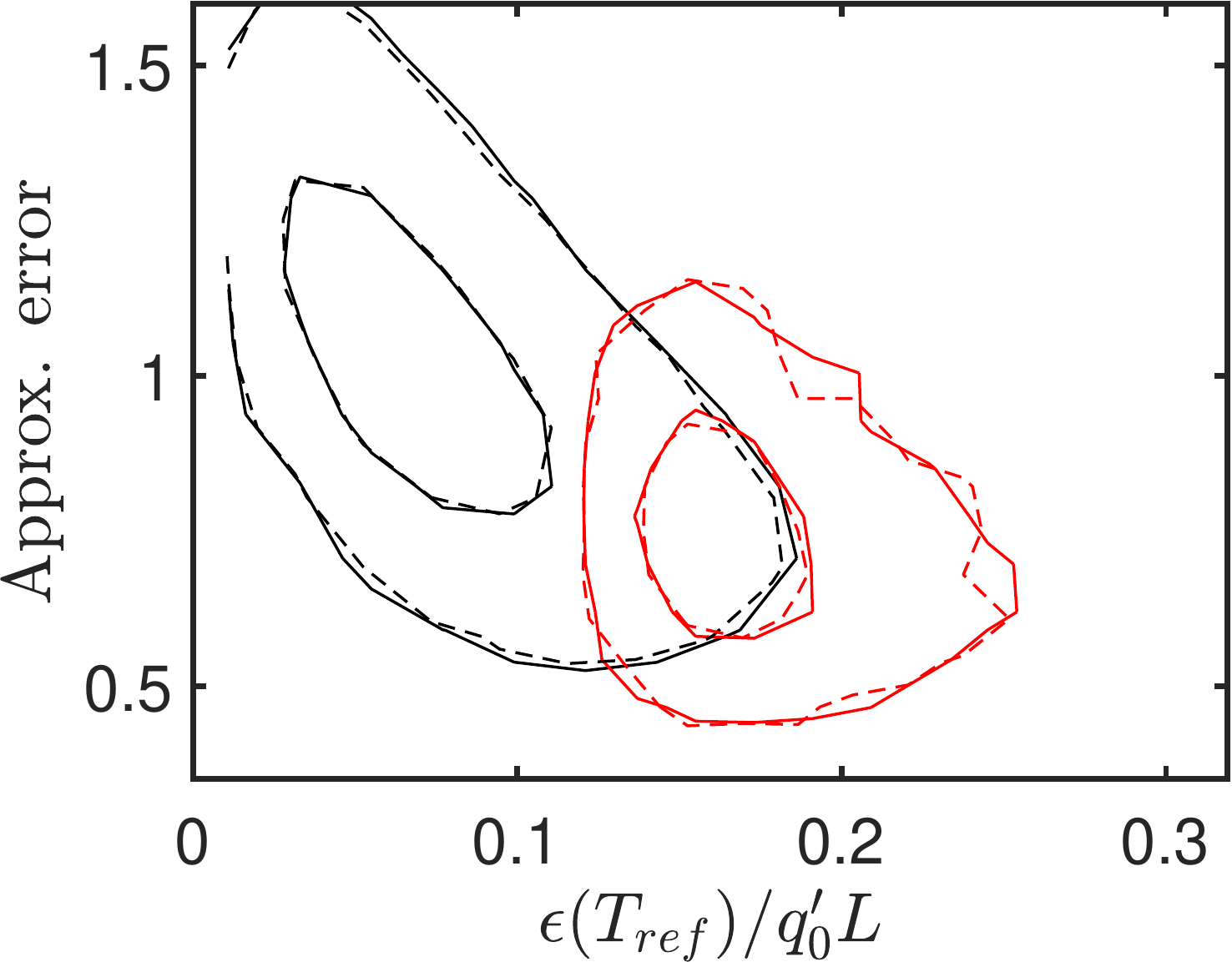}%
\mylab{-.07\textwidth}{.28\textwidth}{(b)}%
}%
\caption{Joint p.d.f. of the relative approximation error, as defined in table
\ref{tab:temperr}, versus the experimental divergence at $\omega'_0 T_{{\mbox{\tiny\it ref}}}
=4.5$. (a) Case 0 in table \ref{tab:caseid}; template is a vortex. (b) Case 10; template is
a dipole.
Black lines use all the experimental cells. Red lines only use the most significant
$n_{keep}=5$ cells in each experiment. 768 experiments, classified using $\|\uvec\|_2$.
Contours contain 50\% and 95\% of the probability. \solid, Tested using the training set;
\dashed, using an independent test set. 
 } 
\la{fig:converse}
\end{figure}

The question of whether the relation between templates and significance can be reversed is
tested in figure \ref{fig:converse}, which shows joint p.d.f.s of the approximation error
between a template and the flow in the neighbourhood of a cell versus the magnitude of the
perturbation resulting from modifying that cell. Contrary to previous figures, the p.d.f.s
drawn in black in figure \ref{fig:converse} are compiled over all the cells of the original
flow, without classification, and it is clear that there is an inverse relation between the
two quantities: cells whose flow neighbourhood is better approximated by a significant
template are also more causally significant. The p.d.f.s drawn in the figure
with red lines only include cells previously classified as most significant. They are the
right-most end of the unconditional distributions. The figure displays separately the result
of using test data from the set used to train the template, and data from a separate set of
experiments, not used in the training. They agree well, decreasing the possibility of
template overfitting.

\section{Vortices and dipoles}\la{sec:physics}

Although a detailed discussion of the dynamics of two-dimensional turbulence is beyond the
scope of the present paper, which is mostly concerned with the methodological aspects of how
to best collect the information required for scientific discovery, it is still useful to
briefly review the physical significance of the collected data.

The most interesting aspect of the above results is the classification of significant
structures into vortices and dipoles.

Individual vortices are widely recognised to be the main coherent structures of
two-dimensional turbulent flows \cite{mcwilliams90b}, where they persist because they
represent droplets of depleted nonlinearity \cite{Push:Bos:2014}. But vortex dipoles
(`modons') have also been investigated as important components of such flows, particularly
in geophysics \cite{Flierl:80,mcwilliams80}, where they retain their individuality for long
times in the atmosphere and in the ocean. They are also relevant in stratified flows, where
they form naturally \cite{Vor:Afa:Fil:91}, and in Bose--Einstein condensates
\cite{NeelyEtal:2010}. However, although the presence of dipoles in non-rotating
two-dimensional turbulence is also well known, they have tended to be considered `transient'
structures, whose importance is usually not emphasised \cite{mcwilliams90b}. The present
results suggest that they deserve a second look. Simulations of a point-vortex gas, which
can be viewed as a non-dissipative model for two-dimensional turbulence \cite{Benzi92}, show
that dipoles form spontaneously and suggest that their relevant property in this context is
that they carry no net circulation, so that their interaction with the rest of the flow is
relatively weak. They are therefore able to move for long distances before being destroyed
by collisions with other objects, and they stir the flow in ways similar to neutral
particles in a plasma.

\begin{figure}
\centerline{%
\includegraphics[height=.32\textwidth,clip]{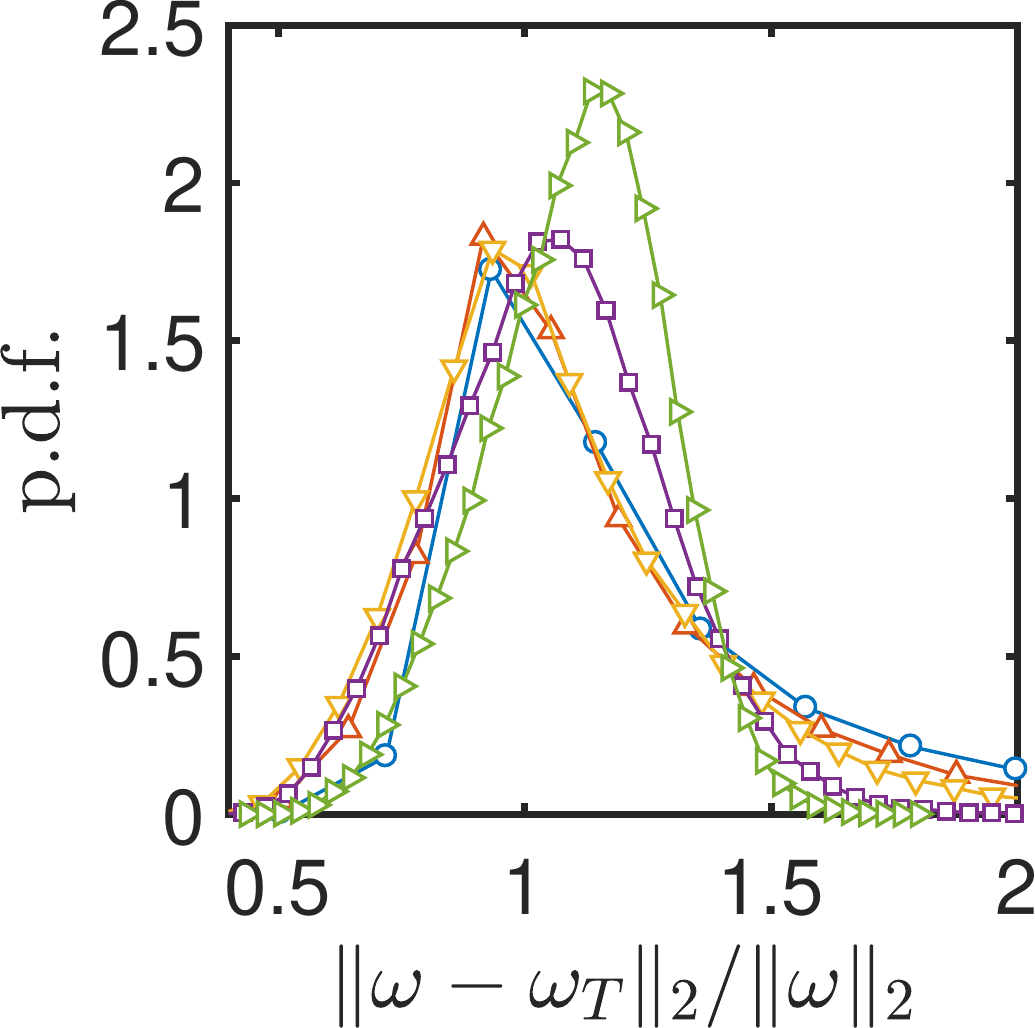}%
\mylab{-.15\textwidth}{.335\textwidth}{(a)}%
\hspace*{2mm}%
\includegraphics[height=.32\textwidth,clip]{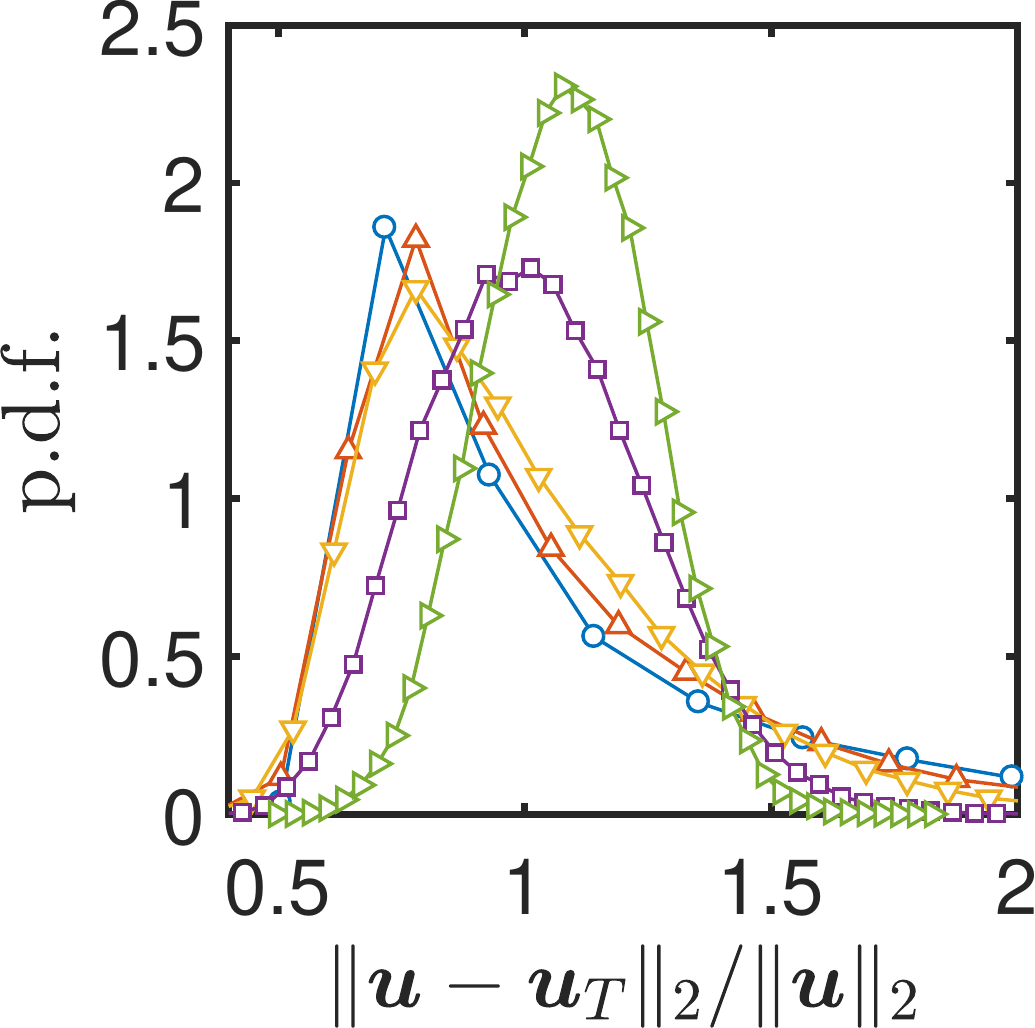}%
\mylab{-.15\textwidth}{.335\textwidth}{(b)}%
\hspace*{2mm}%
\includegraphics[height=.31\textwidth,clip]{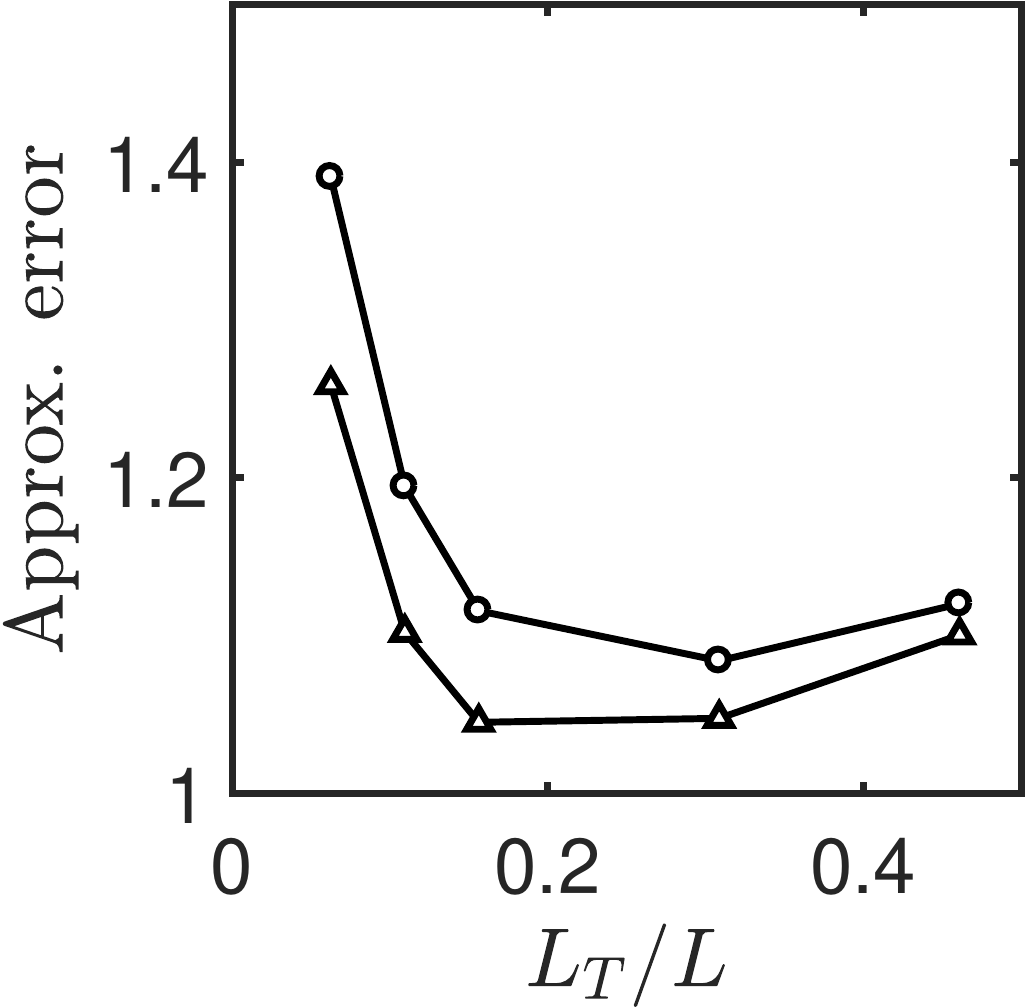}%
\mylab{-.15\textwidth}{.335\textwidth}{(c)}%
}%
\caption{(a,b) P.d.f. of the relative unconditional approximation error between flow and
template, as defined in table \ref{tab:temperr}. Templates are scaled to different sizes,
$L_T$, before being fitted to the flow. \circle, $L_T/L=0.0625$; \trian, 0.101; \dtrian,
0.156; \squar, 0.309 (this is the unscaled template from the $N_c=10$ grid); \rtrian, 0.461.
(a) Template is a vortex.
(b) Template is a dipole.
(c) Average approximation error as a function of $L_T$. \circle, Vortex template; \trian, dipole. 
The template intensity is scaled to match the mean flow intensity.
} 
\la{fig:flow2temp}
\end{figure}

It should be clear that, at this stage of the analysis, the templates obtained above should
be considered models for flow neighbourhoods, independently of how they were initially
obtained, and that we can test how representative of the flow are they without reference to
causality tests. This is done in figure \ref{fig:flow2temp} for vortices and dipoles. In
each case, templates are geometrically scaled to several sizes, since the size at which they
optimally fit the flow is a parameter to be determined, rather than assumed. To keep the
samples statistically equivalent, each template is fitted to flow neighbourhoods centred at
the same $N_c=10$ test grid, and the resulting p.d.f.s are shown in figures
\ref{fig:flow2temp}(a,b). The mean relative fitting error is given in figure
\ref{fig:flow2temp}(c), and should be compared to those in the third column of table
\ref{tab:temperr}, which measures the same quantity between geometrically unscaled templates
and cells classified by their significance. Note that the template size, $L_T$, corresponds
in that case to three test cells, so that $L_T/L=0.309$.

It is interesting that the fitting error in figure \ref{fig:flow2temp}(c) is lowest at
$L_T/L=0.2$--0.4 for both templates, which agrees well with our previous estimates of the
size of the structures from the significance tests. As explained above, the intensity of the
templates has been scaled to match the global intensity of the flow fluctuations, and the
message of figure \ref{fig:flow2temp}(c) is that very small and very large structures of
average intensity do not exist. Recall that the cell size was chosen from the flow spectrum,
and that it was later confirmed, by testing several test partitions, that the vortex and
dipole sizes are comparable to the spectral scale.
 
\begin{figure}
\centerline{%
\raisebox{.06\textwidth}[0mm][0mm]{%
\includegraphics[height=.275\textwidth,clip]{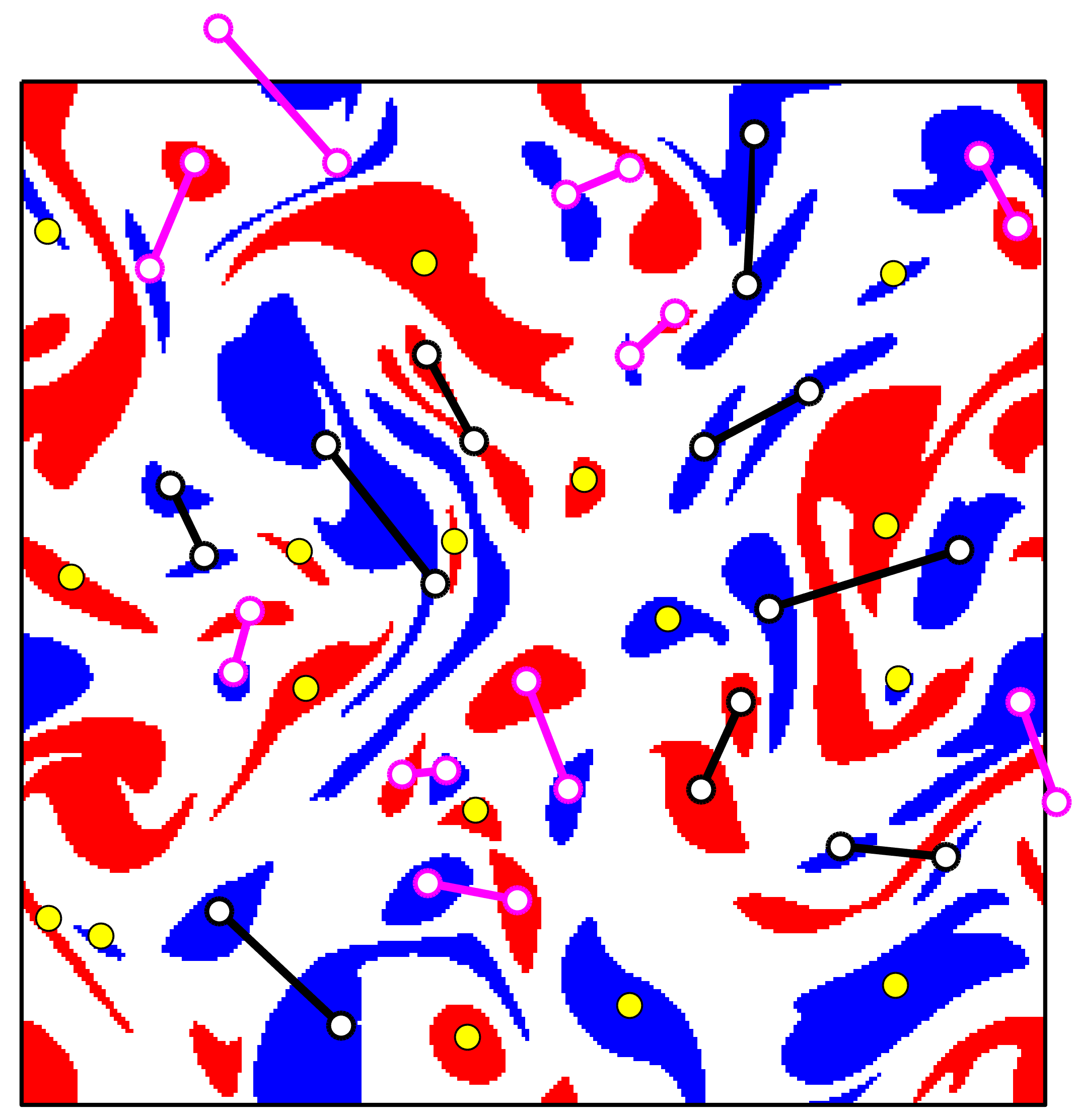}}
\mylab{-.15\textwidth}{.33\textwidth}{(a)}%
\hspace*{3mm}%
\includegraphics[height=.32\textwidth,clip]{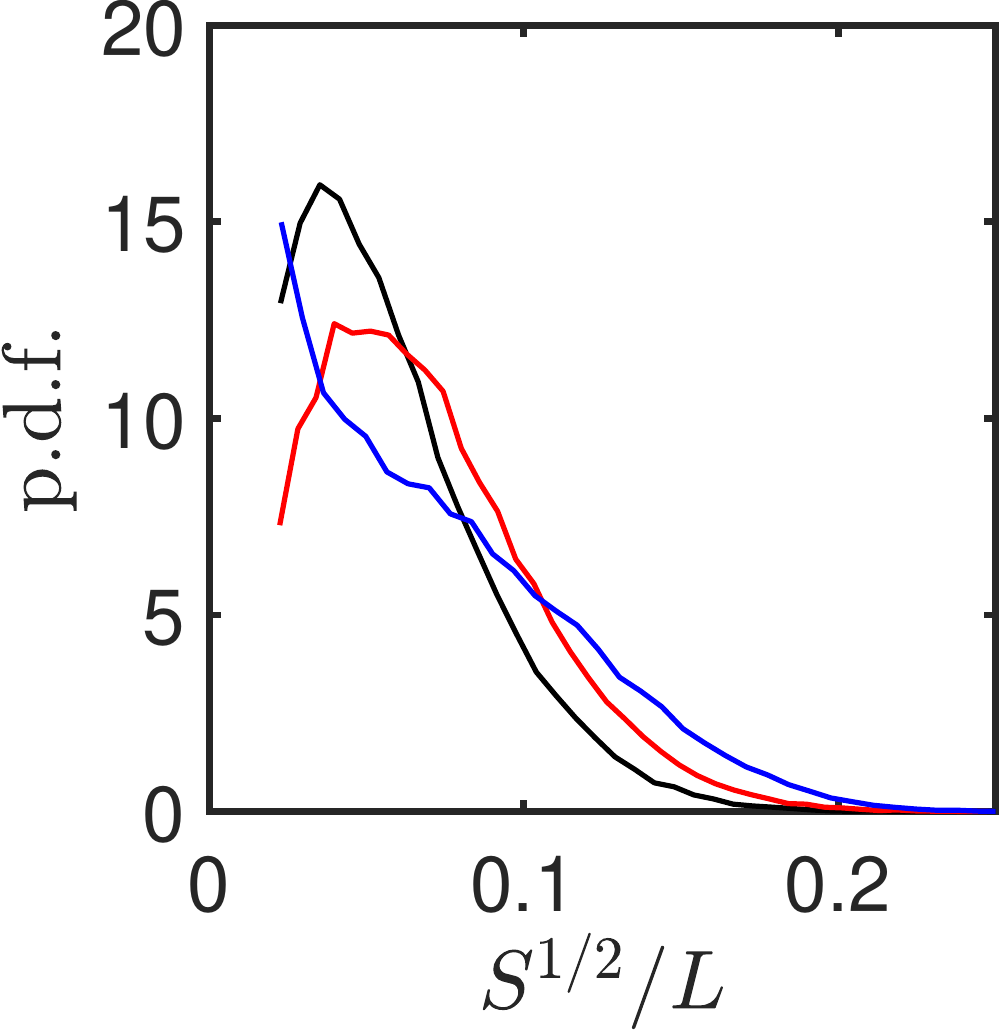}
\mylab{-.15\textwidth}{.33\textwidth}{(b)}%
\hspace*{2mm}%
\includegraphics[height=.32\textwidth,clip]{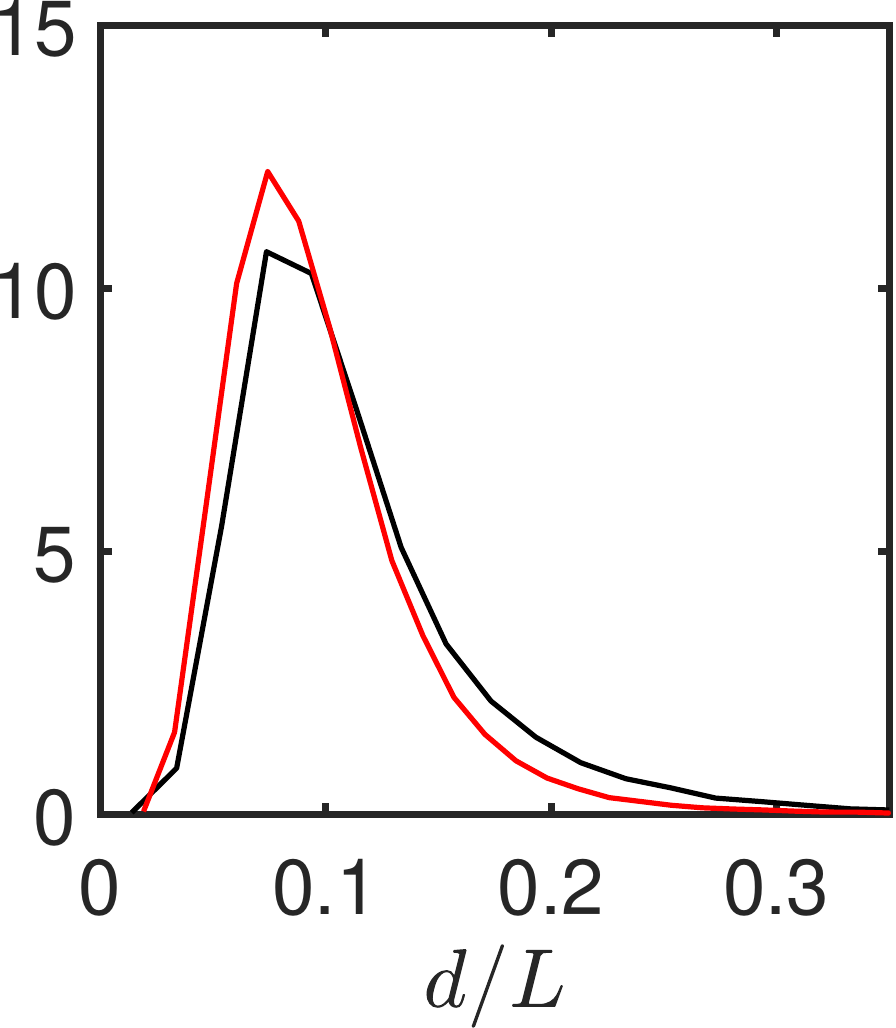}
\mylab{-.15\textwidth}{.33\textwidth}{(c)}%
}%
\caption{(a) Sample segmented field with pair identification. Blue vortices are counterclockwise, and
red ones are clockwise. Black connectors are corrotating pairs, magenta are counterrotating
dipoles, and yellow markers are unpaired vortices. Note that, because of their irregular
shape, some centres of gravity fall outside their vortex. Vorticity threshold
$|\omega|=0.9\omega'_0$.
(b) P.d.f. of the vortex area, $S$. Black, vortices in corrotating pairs; red, counterrotating dipoles; blue, unpaired vortices.
(c) P.d.f. of distance among paired vortices. Symbols as in (b).
} 
\la{fig:segment}
\end{figure}

To gain a better sense of the prevalence of dipoles in two-dimensional turbulence, figure
\ref{fig:segment}(a) shows a segmentation of a typical flow field into individual positive
and negative vortices. Vortices are defined as connected regions in which $|\omega|\ge
H\omega'_0$. For very low thresholds, $H\approx 0$, the flow separates into a few large
objects that fill the whole box but, as $H$ increases, low-vorticity regions fall below the
threshold, and the strong vorticity breaks into more numerous smaller objects. Beyond a
certain threshold, the number of vortices decreases again, and eventually vanishes when no
vorticity satisfies the thresholding condition. The value $H=0.9$ in figure
\ref{fig:segment} is chosen to maximise the number of individual vortices \cite{moisy04}.
The vortices in figure \ref{fig:segment}(a) have been grouped, whenever possible, into co-
and counter-rotating pairs. Two vortices are considered a potential pair if their areas
differ by less than a factor of $m^2$, which is an adjustable parameter. The figure uses
$m=2$, but statistics compiled with $m=1.5$ and $m=3$ show no substantial differences.
Vortices are paired to the closest unpaired neighbour within their area class, and no vortex
can have more than one partner. Some vortices find no suitable partner, and are left
unpaired. Statistics compiled over approximately 8500 independent flow fields are given in
figures \ref{fig:segment}(b,c), and show that the vortex area, $s$, and the distance $d$
between the components of corrotating and counterrotating pairs are very similar. Their
averages, $(s^{1/2}/L\approx 0.07)$ and $(d/L\approx 0.1)$, are of the same order as the
cell size used in the significance studies. A rough measure of the importance of the pairs
is that, out of approximately $5\times 10^5$ vortices, 48\% are paired to form dipoles, 24\%
are in corrotating pairs, and 28\% are isolated. Most vortices in the flow are thus in the
form of pairs, mostly dipoles. The difference between corrotating and counterrotating pairs
is interesting, but it is probably due to the tendency of corrotating pairs to merge into
single cores \cite{meunier05}, while modons are longer-lasting
\cite{Flierl:80,mcwilliams80}.

\section{Discussion and conclusions}\la{sec:conc}

This paper has explored two related but independent ideas in the context of their use in the
process of scientific discovery, with emphasis on turbulence research.
 
The first idea is the well-known distinction between correlation and causation. We have
noted that, because of the difficulty of doing experiments in turbulence (computational or
otherwise), research has tended to centre on correlations. The wealth of data generated by
direct simulations in the 1990's probably contributed to that trend by creating the illusion
that `we knew everything', and that all questions could be answered.

There are two problems with this illusion. The first one is that, although large data sets
contains many answers and, if we believe that turbulence is ergodic, may even contain `all
the answers', the probability of finding them without active researcher participation is
vanishingly small. Consider how difficult it would be to study shock waves by tracking the
expansion of high-density local condensates in ambient air at equilibrium. Dense-enough
condensates are extremely unlikely to form spontaneously, and shocks have to be set up to be
studied with any efficiency. The same is true of many intermittent but significant processes
in turbulence, especially if we aim at controlling the flow by introducing local
out-of-equilibrium perturbations. We noted in the introduction that correlation is the tool
of prediction, while causation is the tool of control, and studying control perturbations
requires experiments that separate causes from later effects.

The second problem is how to choose which experiments to perform. The classical search for
scientific causation is typically hypothesis-driven. The entry point to figure
\ref{alg:method} is the second step S2 (i.e., the researcher thinks of a model, and designs
experiments to test it). While this `hypothesis-driven searches' have obvious theoretical
appeal, they limit creativity. A model has to be conceived before it can be tested, and paradigm
shifts depend on the imagination of the researchers. However, we have noted that faster
experimental and computational methods open the possibility of what we have called Monte-Carlo 
searchers (perhaps describable as `search-driven hypotheses'), in which experiments
are performed randomly, and evaluated a posteriori in the hope that some of them may be
interesting. This is more expensive than the classical procedure, but may be our best hope
of avoiding ingrained prejudice.

We have illustrated these ideas by the application to two-dimensional turbulence in sections
\S\ref{sec:tur2d} and \S\ref{sec:physics}, and we can now reflect on how far the hopes
expressed above have been realised. The first consideration is cost, which we have already
identified as a pacing item for turbulence research. The whole program in this paper took
about a month of computer time in a medium-sized department cluster, and was
programmed by the author in his spare time over a year. Considerably more time was spent
discussing with colleagues what should be done than in actually doing it. Since the effort
was conceived as a proof of principle, the problem was purposely chosen small, but more
interesting problems are within reach. The basic simulation in \S\ref{sec:tur2d} runs in
about 10 core-seconds, but even a three-dimensional $256^3$ simulation of three-dimensional
isotropic turbulence can be run in two minutes in a modern GPU. A program to address
causality in the three-dimensional turbulence cascade, about which considerably less is
known than in two dimensions, can be performed in a modest GPU cluster in a few months
\cite{jimploff20}. The main roadblock would again be to decide what to do.

The second question is whether something of value has been achieved. As mentioned in the
introduction, little was expected from a problem that is usually considered to be
essentially understood. In fact, two previous versions of the present paper
\cite{jimploff18,jimploff20} missed the dipoles completely, and concluded that the
experiments confirmed the classical vortex-gas model of two-dimensional turbulence. The
dipole template in figure \ref{fig:conditional}(b) was a mildly surprising result of further
postprocessing and, if we admit that surprise is one of the defining ingredients of
discovery \cite{schaff:94}, it was a minor discovery, even if we saw in \S\ref{sec:physics} that
dipoles were not completely unknown in two-dimensional flows, and that finding
them was rather an instance of recalling something that had been forgotten than of
discovering something new. This paper is not the place to embark into the development
of a dipole model of two-dimensional turbulence, and it should be obvious to anybody
familiar with turbulence research that many things need to be done before the paper
becomes of independent interest to fluid mechanics. For example, Reynolds number effects
have not been considered, and neither has any experiment in actual control.
But the fact that something unexpected to the author was found without `prompting' is
encouraging for the future of Monte-Carlo methods in problems in which something is
genuinely unknown, including applications outside the realm of turbulence.

The third question is what have we learned about the process of data exploitation. It is
clear that the scientific method in figure \ref{alg:method} can be seen as an optimisation
loop to search for the `best' hypothesis, and it is natural to ask which parts of it can be
automated. Our study in sections \S\ref{sec:tur2d} and \S\ref{sec:physics} was conceived as
an experimental test of how far the automation process could be pushed. Several things can
be concluded. The first one is that step S1 (observations) can be largely outsourced to the
computer, including the reference to `asking questions and planning'. The experiments in
table \ref{tab:caseid} were selected (on purpose) with little thought about their
significance, although it is difficult to say how much they were influenced by the previous
experience of the author. The same can be said about the parts of step S3 that refer to
testing predictions against observations. Simulations of two-dimensional turbulence are by
now routine, and the classification of the results was outsourced to library programs
(figures \ref{fig:svm} and \ref{fig:effic}). An interesting, although not completely
unexpected, outcome of the experience has been the importance of verification and
validation, as illustrated by the examples in sections \S\ref{sec:verify}, \S\ref{sec:testing} and, up to
point, \S\ref{sec:physics}. Unexpected problems are probably features of any project
involving data analysis, but they are especially dangerous in cases, like the present one,
in which the goal is to probe the unknown.

The last point to consider is step S2 in figure \ref{alg:method} (model generation and
evaluation), which is the core of the discovery process. Even here, something was automated.
The templates in figure \ref{fig:conditional} are flow models, and were obtained
automatically. On the other hand, their interpretation in \S\ref{sec:physics} was entirely
manual, and it is difficult to see at the moment how it could be outsourced to a computer.
This is not only because of the need for a level of intelligence somewhat above present
computer software, but because what are we aiming for is not well defined. If the goal of a
scientific project is to find a `good' model, we need to define precisely what we consider a
`good' hypothesis. This naturally leads to the influence of the target audience on the
definition of the research goal \cite{Arrieta:20}. For a fluid mechanician, the ultimate
model of two-dimensional turbulence are the Navier--Stokes equations, and the ultimate cause
of any observation are its initial conditions, presumably traceable to the Big Bang. But we
mentioned in the introduction that we restrict ourselves here to a shorter time horizon
because of some generalised interest in flow control and, while ultimate causality probably
remains a meaningless concept \cite{russ:12}, causality over a given time interval is well
defined. To paraphrase a well-known conundrum, the cause of a chicken over 20 days is an
egg, but the cause over 40 days is another chicken. The choice of a temporal interval
influences the results and makes them less general, but it makes causality questions and
their answers well posed.

The overall conclusion is that human supervision will probably still be required for
hypothesis refinement for some time, but that Monte Carlo can contribute something even
today. Consider again the interpretation of the scientific method as an optimisation loop to
maximise `understanding'. As such, if the goal and parameters are well-defined, they could
conceivably be incorporated into an automatic optimisation procedure (e.g. a neural network).
However, most classical optimisers, including humans, assume local convexity of the cost
function, which is at the root of our misgivings about researcher originality and prejudice.
Monte Carlo is a different way of looking for a maximum, which, in principle, bypasses the
convexity constraint and escapes local maxima by injecting noise in the process
\cite{Deb:Gup:06}. The best use of Monte Carlo science is probably a partially randomised
search and classification step, followed by repeated human fine tuning. The hopeful news in
fluid mechanics is that doing this is now becoming possible.

\section*{Acknowledgement(s)}
This work was supported by the European Research Council under the Coturb grant
ERC-2014.AdG-669505. 



\appendix
\section{Temporal evolution of the perturbation norm}\la{sec:tnorm}

\begin{figure}
\centerline{%
\includegraphics[width=.40\textwidth,clip]{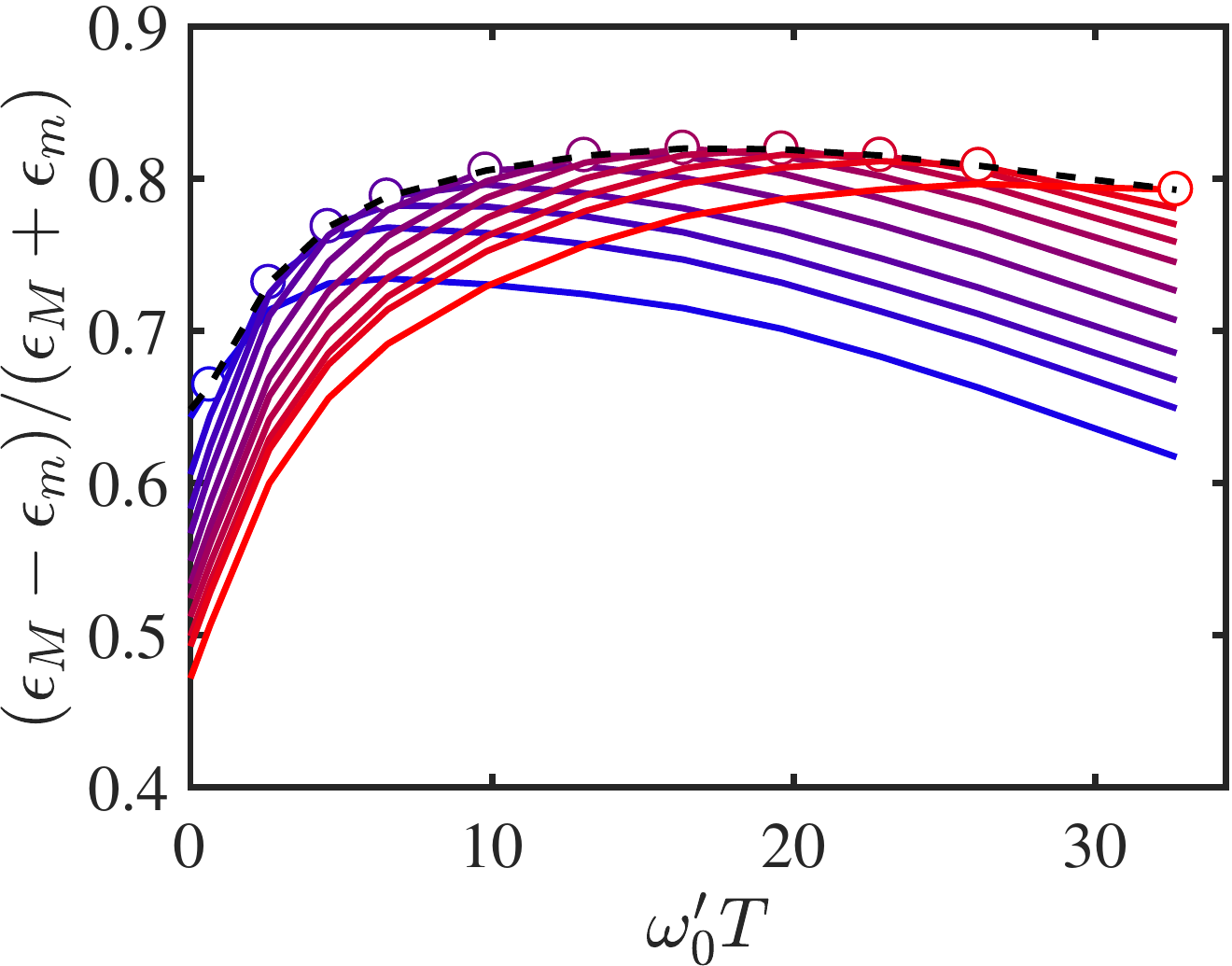}
\mylab{-.05\textwidth}{.27\textwidth}{(a)}%
\hspace*{2mm}%
\includegraphics[width=.40\textwidth,clip]{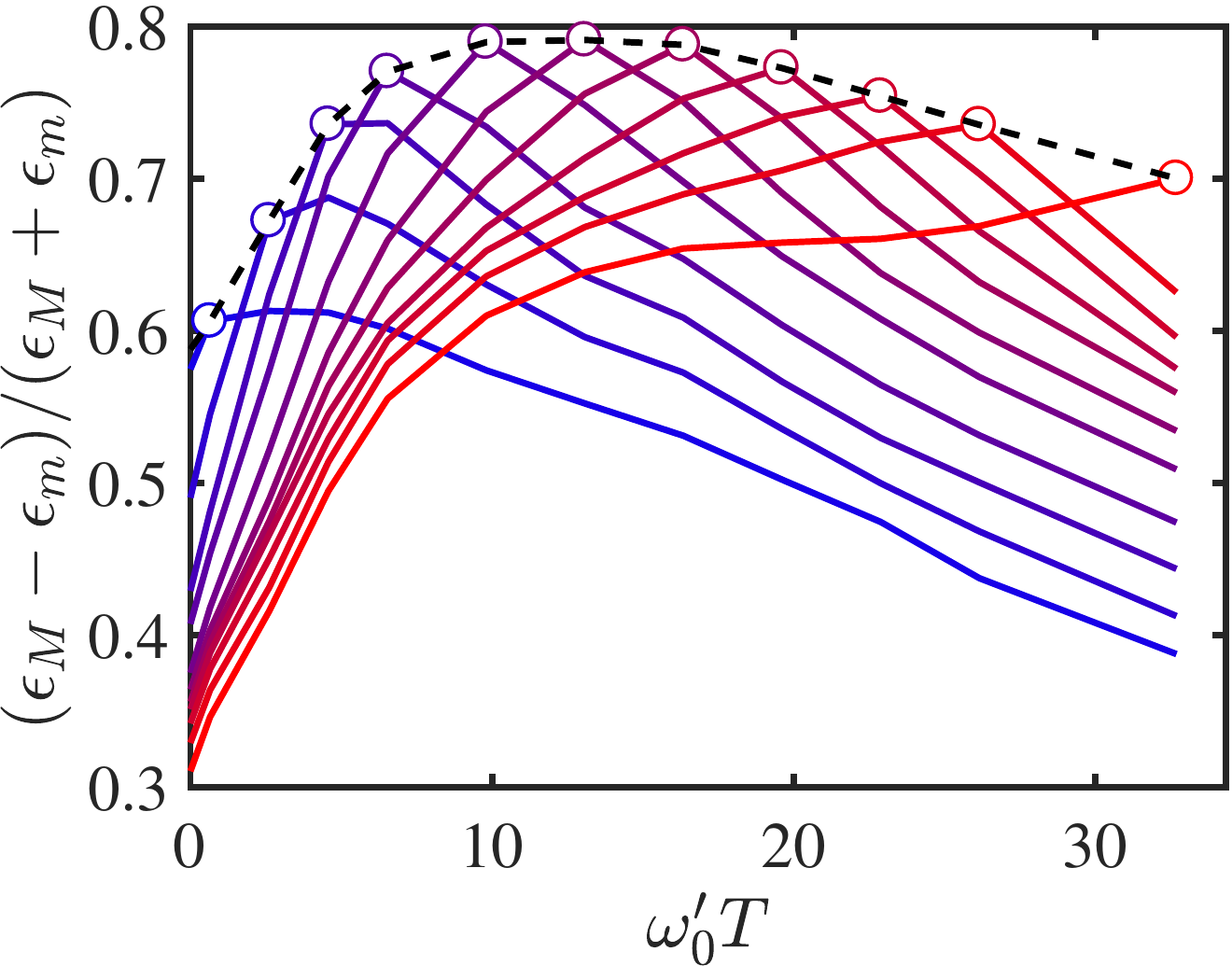}
\mylab{-.05\textwidth}{.27\textwidth}{(b)}%
}%
\vspace{1mm}%
\centerline{%
\includegraphics[width=.40\textwidth,clip]{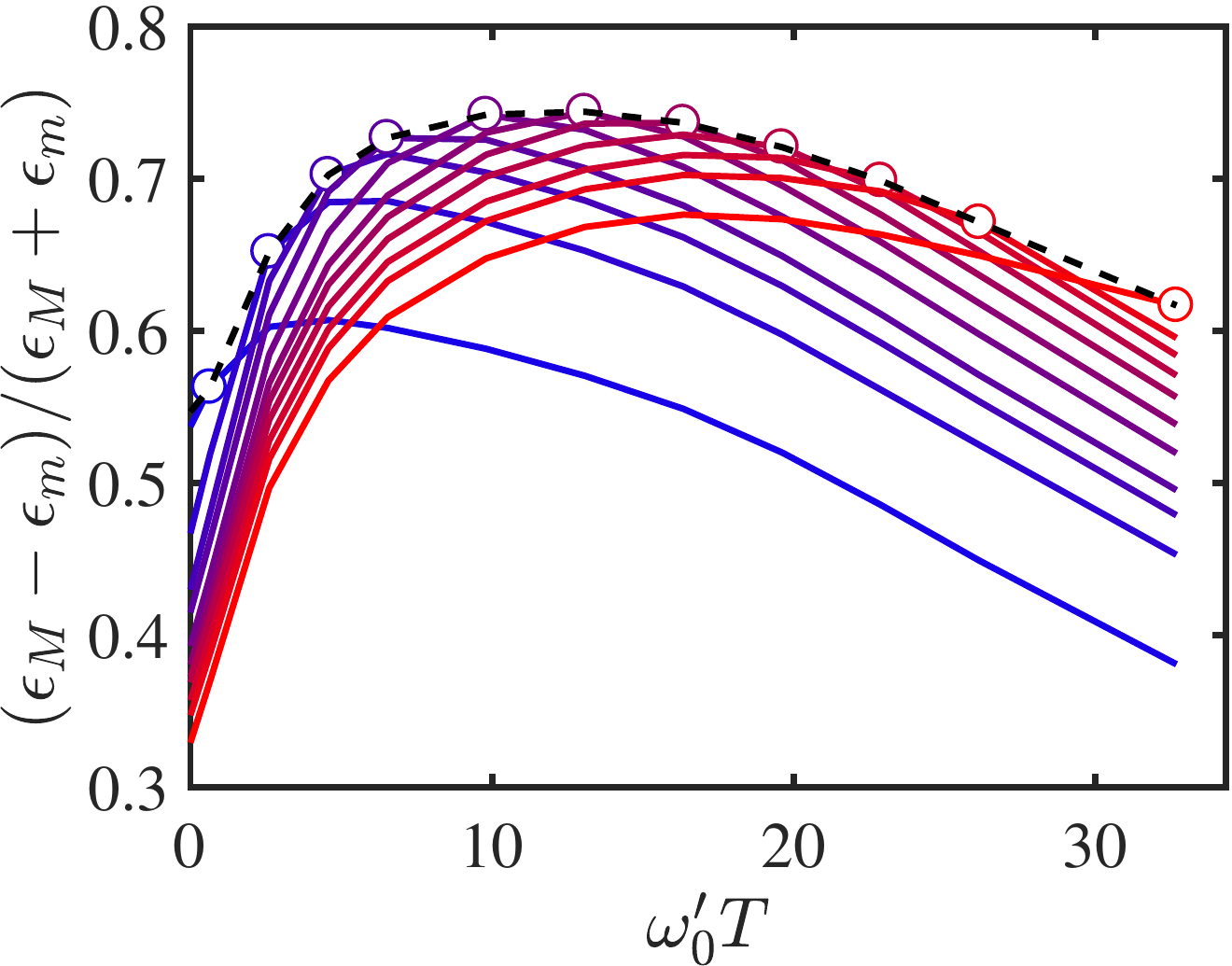}
\mylab{-.05\textwidth}{.27\textwidth}{(c)}%
\hspace*{2mm}%
\includegraphics[width=.40\textwidth,clip]{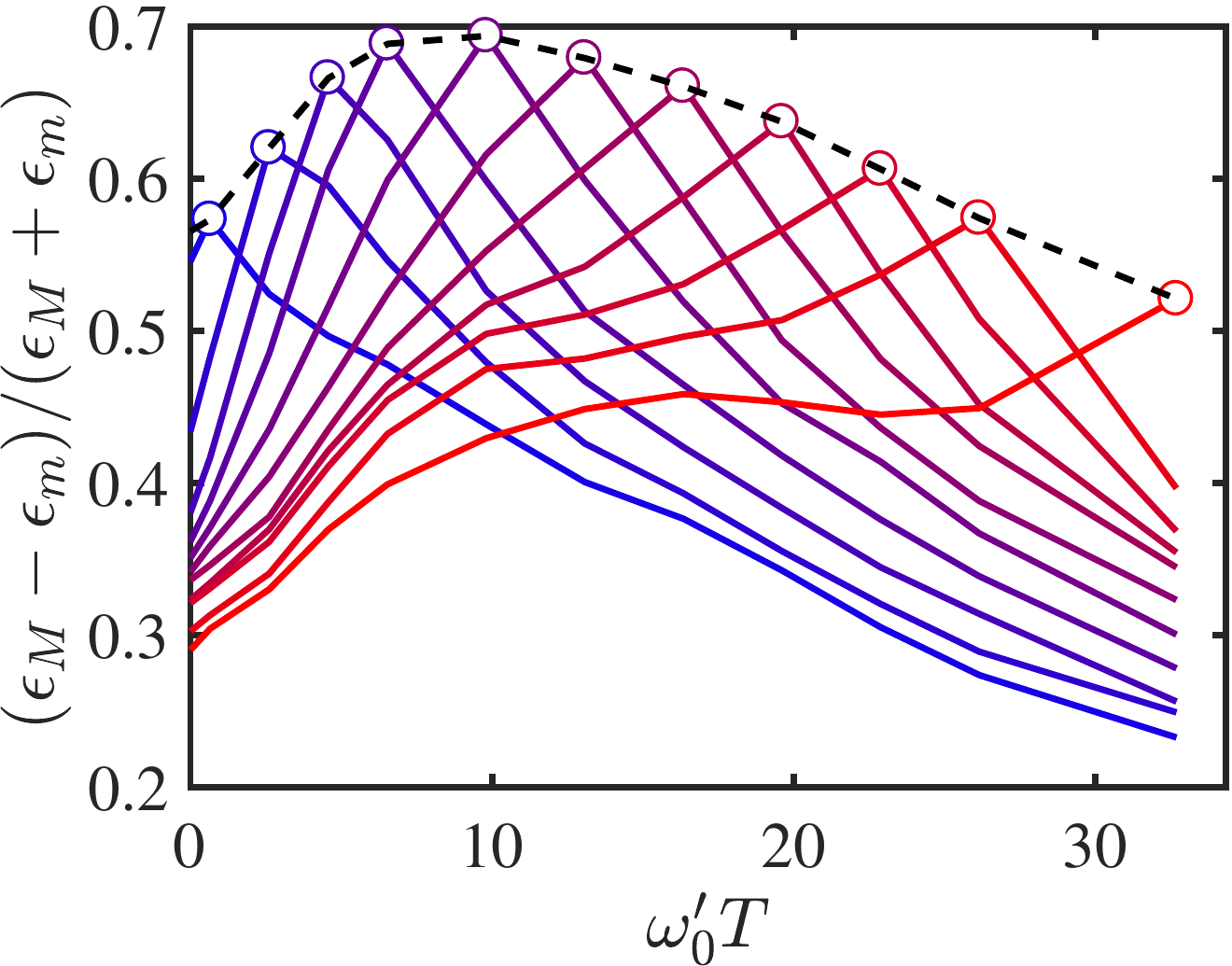}
\mylab{-.05\textwidth}{.27\textwidth}{(d)}%
}%
\caption{Perturbation significance ratio, as a function of time. All panels reflect case 10
in table \ref{tab:caseid}. The circles mark the time at which the classification of disturbances
is made for each curve. The dashed line is the evolvent, and correspond to locally
classified perturbations, as in figure \ref{fig:histime}.
(a) The norm of the perturbations is $\|\uvec\|_2$. (b) $\|\uvec\|_\infty$
(c)  $\|\omega\|_2$. (d) $\|\omega\|_\infty$.
 } 
\la{fig:decay}
\vspace*{3mm}%
\centerline{%
\includegraphics[width=.40\textwidth,clip]{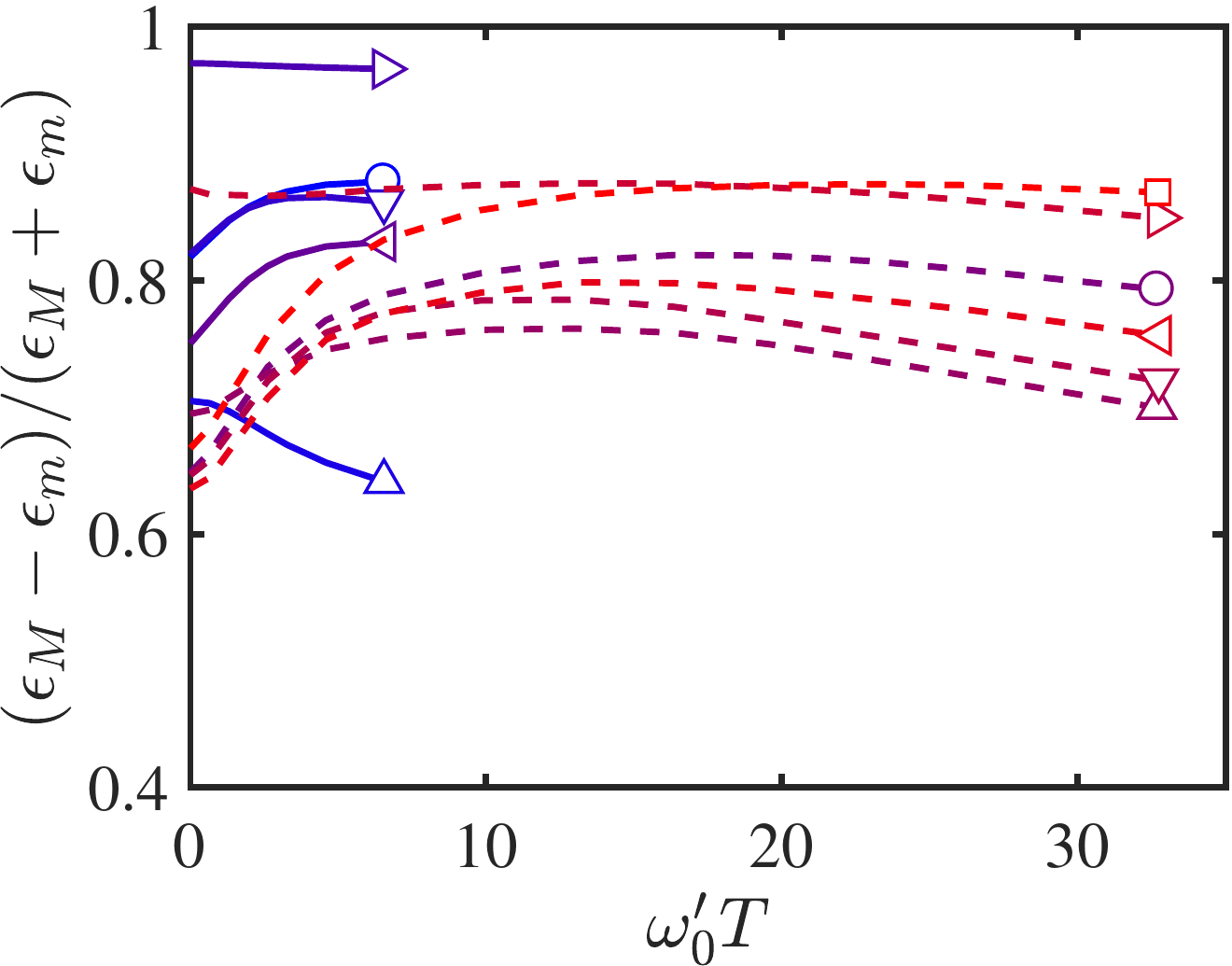}
\mylab{-.07\textwidth}{.27\textwidth}{(a)}%
\hspace*{2mm}%
\includegraphics[width=.40\textwidth,clip]{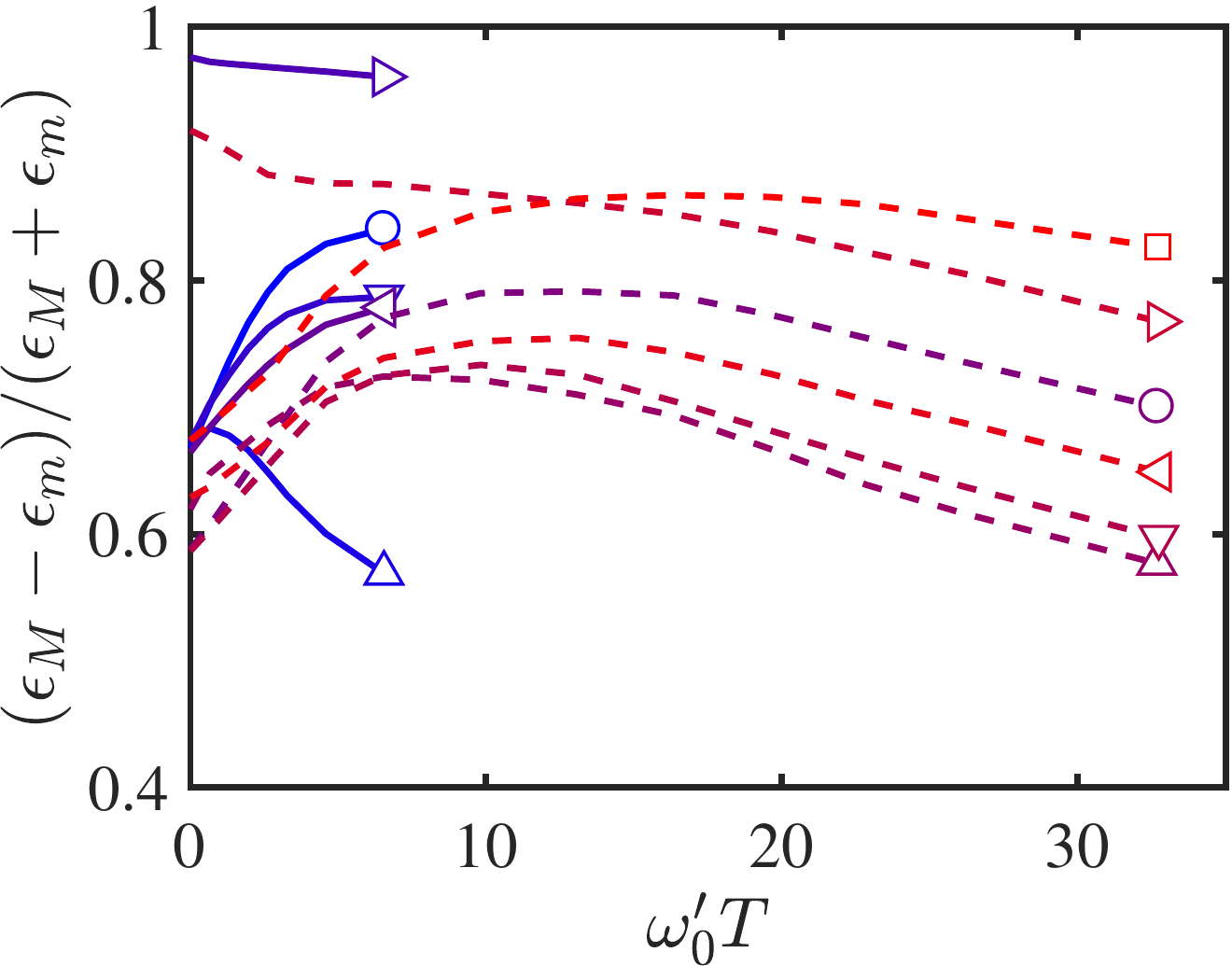}
\mylab{-.07\textwidth}{.27\textwidth}{(b)}%
}%
\vspace{1mm}%
\centerline{%
\includegraphics[width=.40\textwidth,clip]{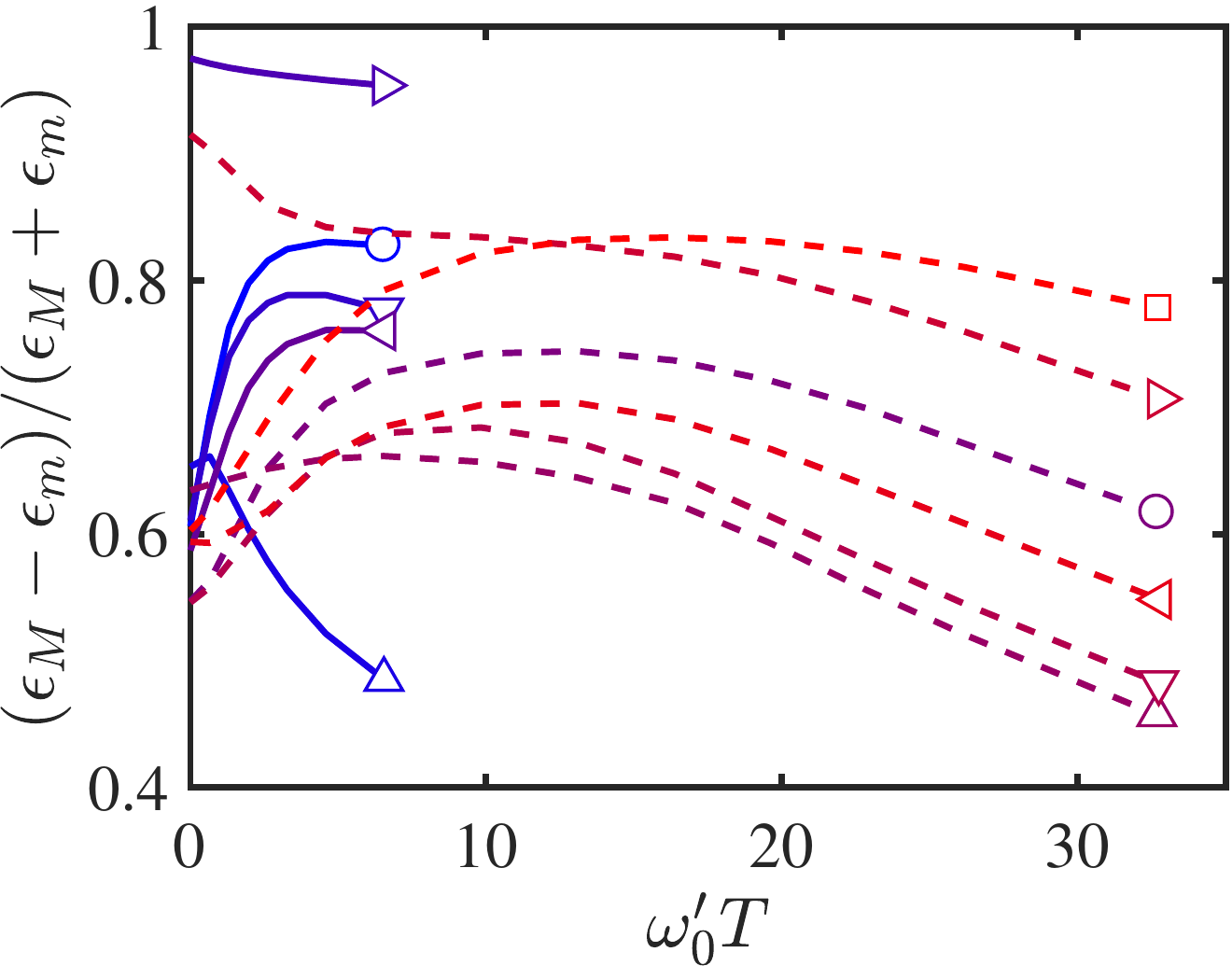}
\mylab{-.07\textwidth}{.27\textwidth}{(c)}%
\hspace*{2mm}%
\includegraphics[width=.40\textwidth,clip]{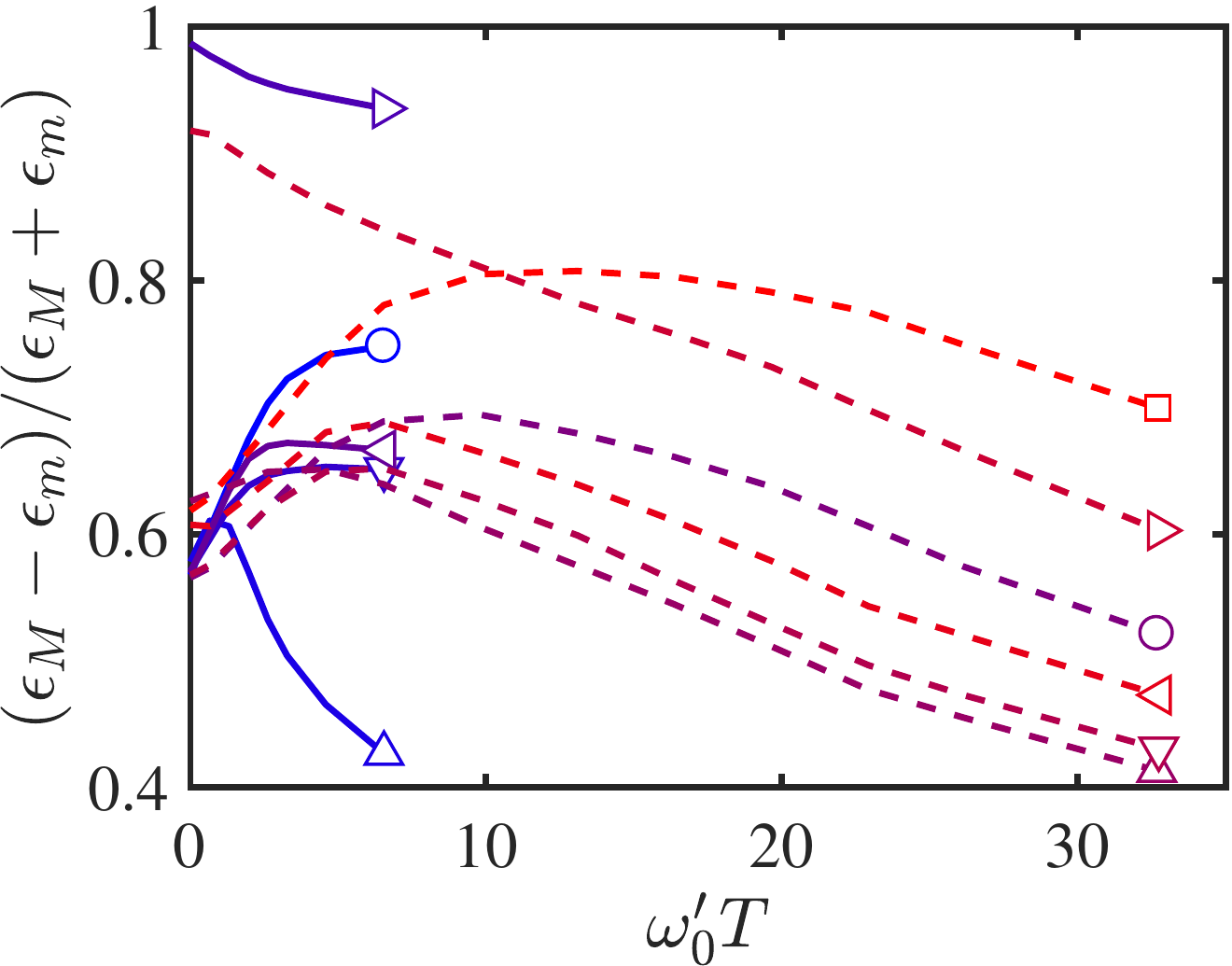}
\mylab{-.07\textwidth}{.27\textwidth}{(d)}%
}%
\caption{Significance ratio, as a function of time, when the significance classes are redefined
at each test time. Symbols as in table \ref{tab:caseid}.
(a) Norm of perturbations is $\|\uvec\|_2$. (b) $\|\uvec\|_\infty$
(c)  $\|\omega\|_2$. (d) $\|\omega\|_\infty$.
 } 
\la{fig:histime}
\end{figure}

An important parameter of the experiments is the time $T_\sref$ at which the significance
classification is performed, and an estimate of an optimum value can be obtained by checking
the evolution of the classification efficiency. A measure of the variability of the effect
of perturbing individual cells is given by the ratio between the largest and smallest
deviations at $T_\sref$, $\epsilon_M$ and $\epsilon_m$, expressed as a relative `efficiency'
\beq
\sigma= \frac{\epsilon_M-\epsilon_m}{\epsilon_M+\epsilon_m},
\la{eq:sigmarel}
\eeq
which varies between $\sigma=0$ for the case in which all perturbations have the same
magnitude, to $\sigma=1$ when the weakest one has zero amplitude. The evolution of this
classification efficiency depends on $T_\sref$. If the separation into significance classes
is done at a given time, and the classification into significance classes is kept constant
thereafter, the efficiency typically peaks at $T_\sref$, and decreases before and after (see
figure \ref{fig:decay}). It is to be expected that, when $\sigma$ is measured in this way,
it decays to zero at very long times, because the flow fields lose the memory of whether
their initial perturbations were more or less influential, and all the experiments tend to
identically distributed random decaying flows $(\epsilon_m=\epsilon_M)$.

A better indication of the perturbation evolution is obtained if cells are reclassified at
each moment, as in figure \ref{fig:histime}, or as in the evolvent in figure
\ref{fig:decay}. Figure \ref{fig:histime} shows that the evolution of $\sigma$ depends on
the perturbation method chosen, but typically peaks at about $\omega_0'T\approx 5-10$, which
probably represents the longest time for which the flow retains the memory of the initial
conditions. The long-time limit is not $\sigma=0$ in this case, but a measure of the
variability among unrelated turbulent flow fields of similar energy, because the distinction
between $\epsilon_m$ and $\epsilon_M$ is constantly maintained by the classifier. As such,
it should presumably be some uniform value, and the slow decay in figure \ref{fig:histime}
represents the evolution of the Reynolds number in these decaying flows. More interesting is the
fast rise, in most experiments, over the initial $\omega'_0T\approx 5$, which suggests that
the initial perturbations need some time to become organised enough to have an effect on the
overall norm. It is interesting that some perturbations (cases 1, 4 and 14 in table
\ref{tab:caseid}) are created `fully formed', and that their significance decays from the
beginning.

Figures \ref{fig:decay} and \ref{fig:histime} include results with the four norms used in
this work. There are substantial differences. As could be expected, $\|\cdot\|_\infty$, which is a
local norm, tends to vary more than the integral $\|\cdot\|_2$, but there are also differences
between the norms based on the velocity and those based on the vorticity. Independently of
the perturbation scheme, the vorticity norms vary more, and their significance ratio grows and decays
faster than those of the velocity.

\section{Vector perturbations}\la{sec:optcomp}

Modifying a vector quantity, such as the velocity, introduces more degrees of freedom than a
scalar one, such as the vorticity in two dimensions. In three-dimensional flows, most
quantities are vectors, and it is important to reduce the degrees of freedom by estimating
which component of the vector would have a greater effect after a given time. Thus, case 10
in table \ref{tab:caseid} does not consist of zeroing the full velocity vector in the cell,
but one of its components, and it is desirable to determine which component to modify for
maximum effect.

Assume that we want to modify a column vector property, $\uvec=\{u_1,u_2\}$, at time $t=0$ by some
amount $\vdelta=\{\delta_1,\delta_2\}$, and that we would like to choose the direction of
$\vdelta$ so that the effect at time $t=T$ is maximum. We will assume that,
although the relation between $\uvec(0)$ and $\uvec(T)$ is not linear, the effect of
mixing several perturbations can be linearised, so that the perturbation created by
$\vdelta(0) = \alpha_1 \vdelta^{(1)}(0) + \alpha_2 \vdelta^{(2)}(0)$ can be approximated as
$\vdelta(T) = \alpha_1 \vdelta^{(1)}(T) + \alpha_2 \vdelta^{(2)}(T)$. This is plausible,
because $\vdelta$ is assumed to affect only a small part of the flow. It will be
numerically tested below. The goal is to choose the $\alpha_j$'s to maximise the quadratic
norm $\|\vdelta(T)\|^2 = (\vdelta^*(T)| \vdelta(T))$, where the asterisk denotes Hermitian
transpose and $(\,|\,)$ is a scalar product. We impose scale by requiring that
\beq
\|\valpha\|^2 = \alpha_1^2+\alpha_2^2=1.
\la{eq:alphanorm}
\eeq
where $\valpha=\{\alpha_1, \alpha_2\}$ is the column vector of coefficients. Assuming linear
superposition of the perturbations, it is easy to check that
\beq
\|\vdelta(T)\|^2 =\valpha^* \matA(T) \valpha, 
\la{eq:alphaJ0}
\eeq
where
\beq
\matA = \left( \begin{array}{cc}
      (\vdelta^{(1)*}|\vdelta^{(1)}) &  (\vdelta^{(1)*}|\vdelta^{(2)}) \\       
      (\vdelta^{(2)*}|\vdelta^{(1)}) &  (\vdelta^{(2)*}|\vdelta^{(2)}) 
      \end{array}\right)
\la{eq:corrmat}
\eeq
is the cross-correlation matrix at time $T$, and  the quantity to be maximised is
\beq
J(\valpha)=\valpha^* \left[\matA(T) - \lambda \matI\right]\valpha, 
\la{eq:alphaJ}
\eeq
where $\matI$ is the identity matrix, and $\lambda$ is a Lagrange multiplier to enforce
\r{eq:alphanorm}. Since $\matA$ is Hermitian, all its eigenvalues are real, and the maximum
of $\|\vdelta(T)\|$ is attained by choosing $\valpha$ as the eigenvector of the largest
eigenvalue. The procedure can easily be generalised to more dimensions.

In practice, when manipulating the velocity in the code, the flow is evolved twice for a
short time, $\omega'_0 T_l = O(1)$, using each time linearised initial perturbations for one
velocity component. For example, if the unperturbed initial condition is $\uvec_0(0)$, and
the initial perturbation resulting from modifying only the $u_1$ component by some
(non-infinitesimal) procedure is $\bvec_1$, the code is first run with the `linearisable'
initial condition $\uvec^{(1)}(0)=\uvec_0(0) + \eta \bvec_1$, where, typically,
$\eta=10^{-2}$. The process is repeated with perturbations to the other velocity component. The
matrix $\matA$ is formed from the observed perturbations $\vdelta^{(j)}(T_l)
=\uvec^{(j)}(T_l) -\uvec_0(T_l)$, and the final `optimal' initial condition is chosen to be
\beq
\uvec_{opt}=(1-\alpha^{(1)}_1-\alpha^{(1)}_2) \uvec_0(0) + 
\alpha^{(1)}_1 \bvec_1(0) + \alpha^{(1)}_2 \bvec_2(0),
\la{eq:uopt}
\eeq
where $\valpha^{(1)}$ is the eigenvector of the largest eigenvalue, $\lambda^{(1)}$. Note
that the linearising factor, $\eta$, has been removed from the final perturbation.

\begin{figure}
\centerline{%
\includegraphics[width=.32\textwidth,clip]{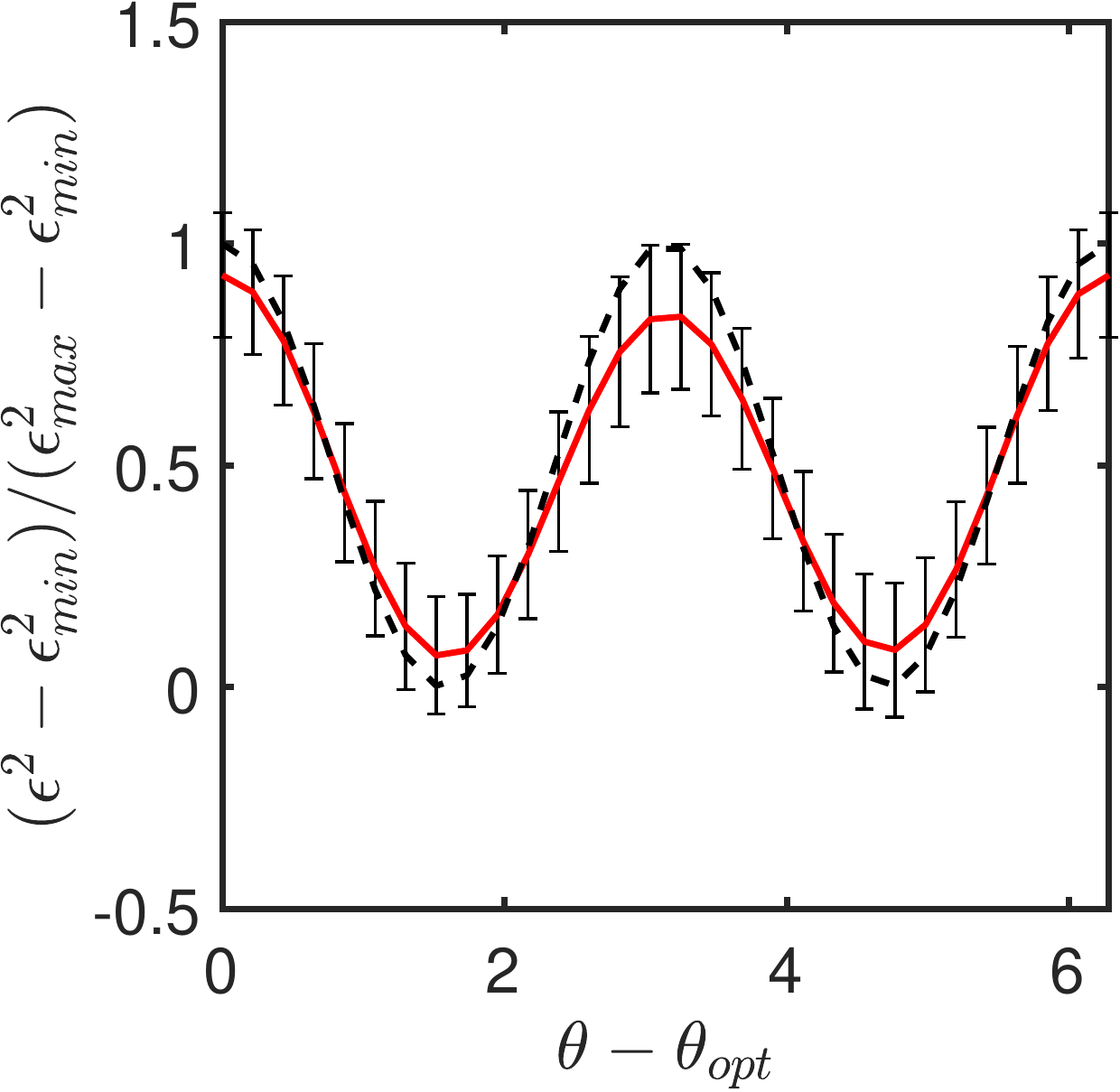}
\mylab{-.16\textwidth}{.32\textwidth}{(a)}%
\hspace*{2mm}%
\includegraphics[width=.32\textwidth,clip]{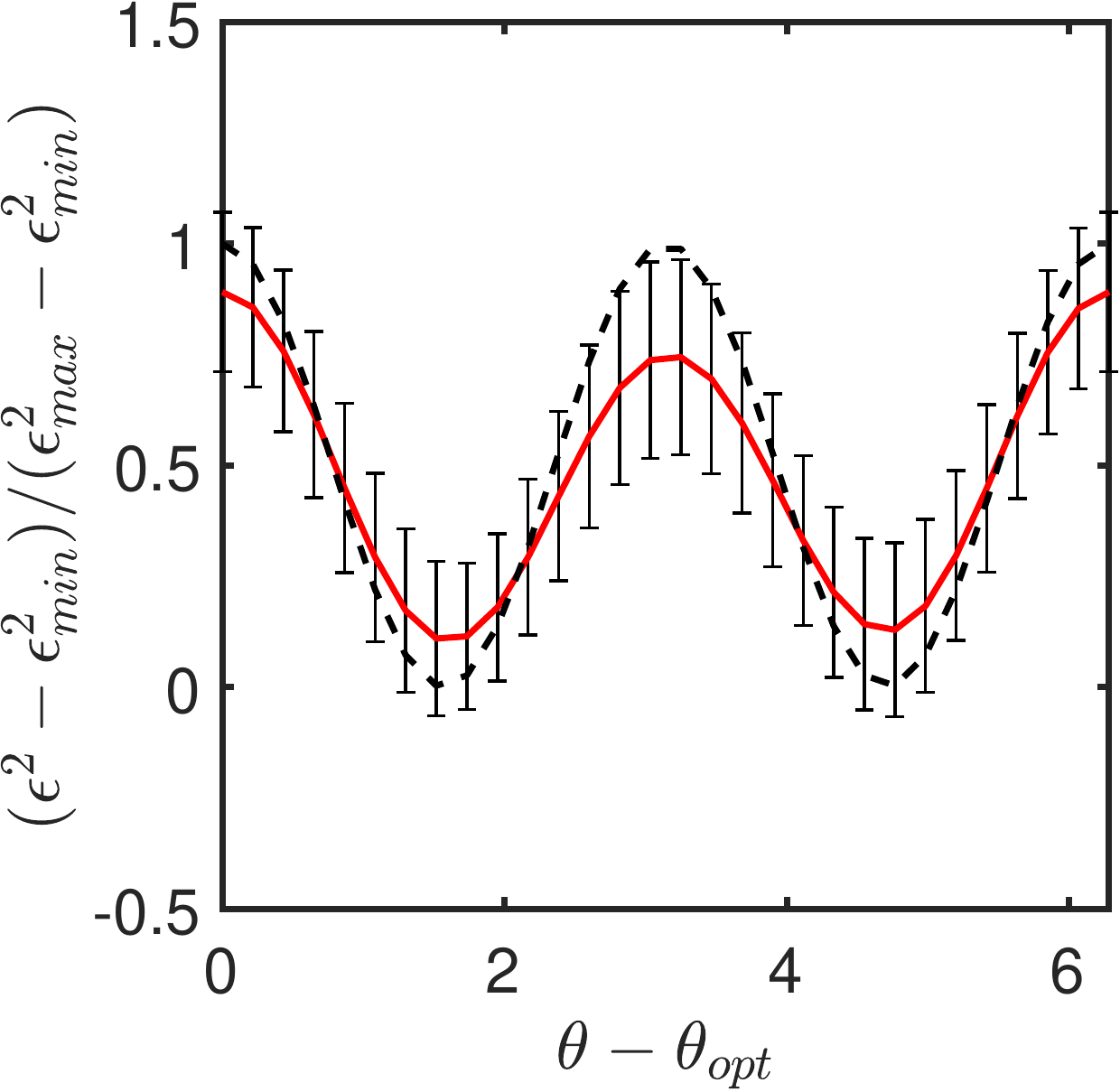}
\mylab{-.16\textwidth}{.32\textwidth}{(b)}%
\hspace*{2mm}%
\includegraphics[width=.32\textwidth,clip]{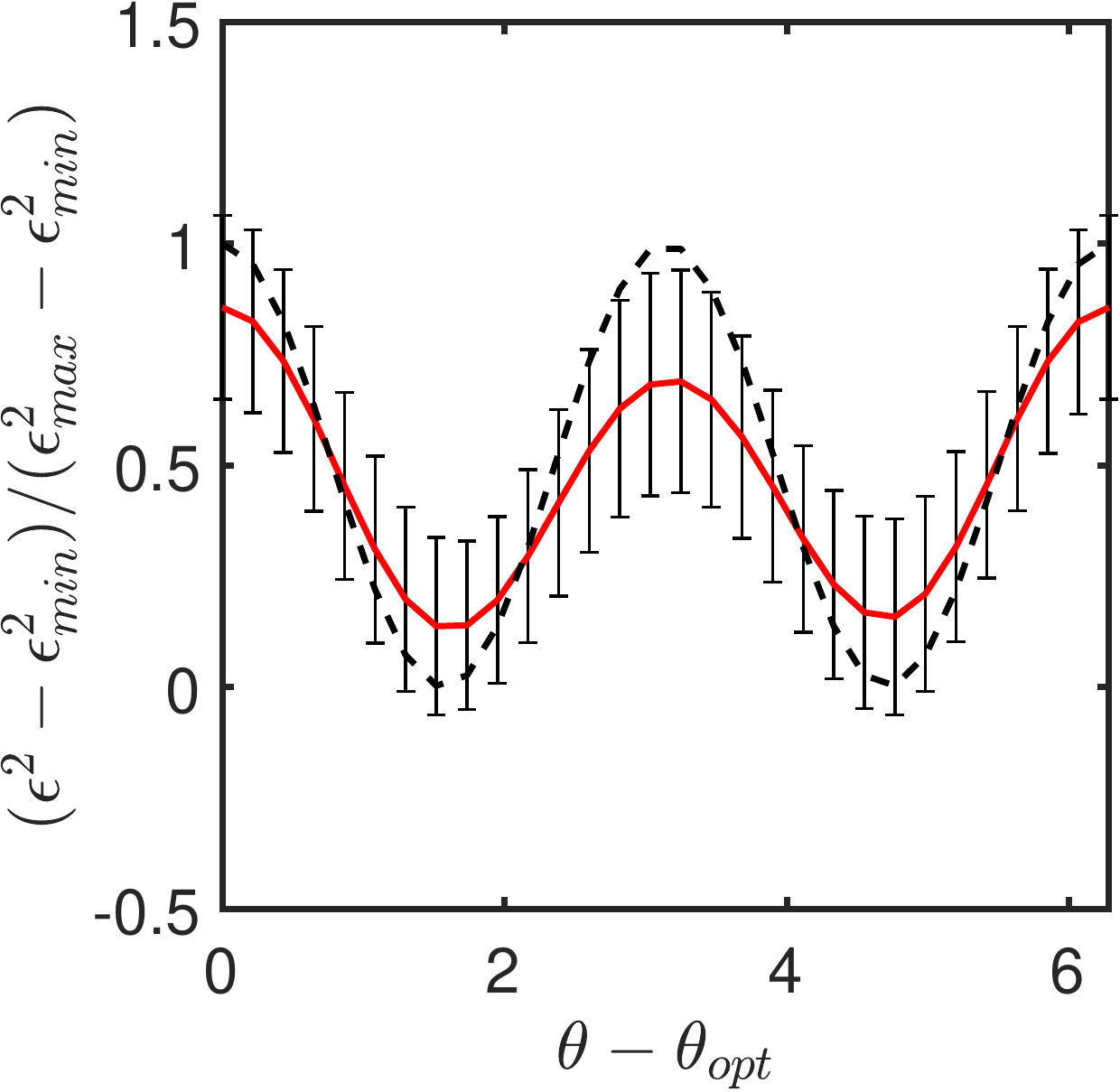}
\mylab{-.16\textwidth}{.32\textwidth}{(c)}%
}%
\caption{Solid line is the actual $\epsilon=\|u\|_2$ perturbation intensity at time $T$, as
a function of the initial orientation $\theta-\theta_{opt}$ with respect to the linearly
determined optimum. Error bars are $\pm 1$ standard deviation. Dashed line is \r{eq:sincos}.
(a) $\omega'_0T= \omega'_0T_l= 1.3$. (b) $\omega'_0T=2.6$. (c) $\omega'_0T=4.5$. Case 10,
averaged over 48 different initial fields.
} 
\la{fig:testlin}
\end{figure}

In the two-dimensional case, we can express $\valpha= \cos(\theta) \valpha^{(1)} +
\sin(\theta) \valpha^{(2)}$. If the initial perturbation is chosen in the direction of a
generic $\valpha$, instead of parallel to the linearised optimal eigenvector, the linear
prediction is that,
\beq
\|\vdelta(T)\|^2 = \cos^2(\theta) \|\vdelta(T)\|^2_{max} +\sin^2(\theta)  \|\vdelta(T)\|^2_{min} , 
\la{eq:sincos}
\eeq
where the maximum and minimum are taken over $\theta$.
This is tested in figure \ref{fig:testlin}, which shows the degradation of linearity as a
function of time. In figure \ref{fig:testlin}(a), which is drawn at the linearisation
time $T_l$, the experimental results follow reasonably closely the theoretical prediction
\r{eq:sincos}, but the agreement gets worse as time progresses. Because of the way the
different experiments are normalised in the figure, each of them oscillates between zero and
one, but the average does not exactly follow \r{eq:sincos} because linear mixing
does not exactly hold for nonlinear perturbations, and the optimal
orientation does not exactly agree with the linearised prediction. In the limit of zero
agreement, the experimental average would tend to a uniform value of 0.5, but it is encouraging
that  in figure \ref{fig:testlin}(c), which corresponds to the two rightmost flow fields in
figure \ref{fig:perevol} in the text, and which can be considered as fully nonlinear, the maximum
averaged perturbation is still at the theoretical linear orientation, and the standard
deviation is still moderate.
   
\end{document}